\documentclass[prl,preprint]{revtex4}
\usepackage{graphicx}

\begin{document}

\title{Stochastic dynamics of a finite-size spiking neural network\footnote{Neural Computation, in press}}
\author{H\'edi Soula}
\author{Carson C. Chow}
\affiliation{Laboratory of Biological Modeling,
NIDDK, NIH, 
Bethesda, MD 20892}

\begin{abstract}
We present a simple Markov model of spiking neural dynamics that
  can be analytically solved to  characterize the stochastic dynamics of a 
finite-size spiking neural network.  We give closed-form estimates for the
equilibrium distribution, mean rate, variance and autocorrelation
function of the network
activity.  The model is applicable to any network where the
probability of firing of a neuron in the network only depends on the
number of neurons that fired in a previous temporal epoch.   Networks
with statistically homogeneous connectivity and membrane and
synaptic time constants that are not excessively long could satisfy
these  conditions.  Our model completely accounts for  
the size of the network and correlations in the firing activity. It also
allows us to examine how the network dynamics can deviate from mean-field theory.
We show that the model and solutions are applicable to spiking neural networks in
biophysically plausible parameter regimes.
\end{abstract}

\maketitle

\section{Introduction}

Neurons in the cortex, while exhibiting signs of synchrony in certain
states and tasks~\cite{singer1995,steinmetz2000,fries2001,pesaran2002}, mostly fire
stochastically or asynchronously~\cite{softky1993}. 
Previous theoretical and computational work on the stochastic dynamics
of neuronal networks have mostly focused 
on the behavior of networks in the infinite size limit with and
without the presence of external noise.  The bulk of these studies utilize
a mean-field theory approach, which presumes
self-consistency between the input to a given neuron from the
collective dynamics of the network with the output of that
neuron and either ignores fluctuations or assumes that
the fluctuations obey a prescribed statistical distribution, e.g.
\cite{amit1997a,amit1997b,vanvreeswijk1996,vanvreeswijk1998,gerstner2000,cai2004}
and \cite{gerstner2002a} for a review. Within the mean-field
framework, the statistical 
distribution of fluctuations may be directly computed using a Fokker-Planck approach \cite{abbott1993,treves1993,fusi1999,golomb2000,brunel2000a,nykamp2000,hansel2003,delgiudice2003,brunel2006} or estimated from the response of one individual neuron submitted to noise \cite{plesser2000,salinas2002,fourcaud2002,soula2006}.

%For example, in the case of integrate-and-fire neurons, the resulting dynamics and phase transitions have been analyzed for networks with sparse connections \cite{}, inhibitory connections \cite{}, stochastic networks \cite{delgiudice2003} and for linear (VLSI) neurons \cite{fusi1999}. 

Although, correlations in mean-field or near mean-field networks can
be nontrivial~\cite{ginzburg1994,vanvreeswijk1996,vanvreeswijk1998,meyer2002}, in general,
mean-field theory is strictly valid only if the size of the network is
large enough, the connections are sparse enough or the external noise
is large  
enough to decorrelate neurons or suppress fluctuations.   Hence,
mean-field theory may not capture correlations in the firing activity of the network
that could be important for the transmission and coding of information. 

It is well known that finite-size effects can contribute to
fluctuations and correlations~\cite{brunel2003b}.
It could be possible that for some areas of the brain, small
regions are statistically homogeneous in that the probability
of firing of a given neuron is mostly affected by a common input and the neural activity within a given neighborhood.  These neighborhoods may
then be influenced by finite-size effects.     
However,  the effect of finite size on neural
circuit dynamics is not fully understood. Finite-size effects have been
considered previously using expansions around 
mean-field theory~\cite{ginzburg1994,meyer2002,mattia2002}. 
It would be useful to develop analytical methods that could
account for correlated fluctuations due to the finite size of the
network away from the mean-field limit.

In general, finite-size effects are difficult to analyze. We circumvent
some of the difficulties by considering a simple Markov model where the
firing activity of a neuron at a given temporal epoch only depends on
the activity of the network at a 
previous epoch.  This simplification, which presumes statistical homogeneity in the network,  allows us to produce
analytical expressions for the equilibrium distribution, mean,
variance, and autocorrelation 
function of the network
activity (firing rate) for arbitrary network size.  We find that mean-field theory
can be used to estimate the mean activity but not the variance and
autocorrelation function.
%In addition, we can also capture transient activity and the approach
%to equilbrium. 
Our model can describe the stochastic dynamics of spiking neural networks in biophysically
reasonable parameter regimes.
We apply our formalism to  three different
spiking neuronal models.

\section{The model}

We consider  a simple Markov model of spiking neurons. We assume that the
{\em network activity}, which we define as the number of neurons firing at a given
time, is characterized entirely by the network
activity in a previous temporal epoch.  In particular, the number of neurons
that fire between $t$ and $t+\Delta t$ only depends on the number that
fired between $t -\Delta t$ and $t$.    This amounts to assuming that
the neurons are statistically homogeneous in that the probability that
any neuron will fire is uniform for all neurons in the network
and have a short memory of earlier network activity.  Statistical
homogeneity could arise for example, if the network receives common
input and the recurrent connectivity is effectively homogeneous.  We note that our
model does not necessarily require that the network architecture be
homogeneous, only that the firing statistics of each neuron be
homogeneous.  As we will show later, the model can, in some
circumstances, tolerate random connection heterogeneity. 
Short neuron memory can arise if the membrane and synaptic time
constants are not excessively long.
 
The crucial element for our formalism is the gain or response function of a neuron
$p(i,t)$, which gives the probability of firing of one neuron within
$t$ and $t+\Delta t$ given that $i$ neurons have fired in the previous
epoch. The time dependence can reflect the time dependence of an external input
or a network process such as adaptation or synaptic depression.
Without loss of generality, we can rescale time so that $\Delta t=1$.  

Assuming $p(i,t)$ is the same for all neurons, then the probability
that exactly $j$ neurons in the network will fire between $t$ and $t+1$
given that $i$ neurons fired between $t-1$ and $t$ is  
\begin{equation}
p(X(t+1)=j | X(t)=i) = C_{N}^j p(i,t)^j(1-p(i,t))^{N-j},
\label{prob}
\end{equation}
where $X(t)$ is the number of neurons firing between $t-1$ and $t$ and
$C_{N}^k$ is the binomial coefficient.  
Equation (\ref{prob}) describes a Markov process on the finite set
$\{0,..,N\}$ where $N$ is the maximum number of neurons allowed to
fire on a time interval $\Delta t$. The process can be re-expressed
in terms of a time dependent probability density function (PDF) of the
network activity $P(t)$ and a Markov transition 
matrix (MTM) defined by 
\begin{equation}
M_{ij}(t)\equiv p(X(t+1)=j | X(t)=i) =  C_{N}^k p(i,t)^j(1-p(i,t))^{N-j}.
\label{M}
\end{equation}
$P(t)$ is a row vector on the space $ [0,1]^N$
that obeys 
\begin{equation}
P(t+1)=P(t)M(t)
\label{eq:dyn1}
\end{equation}
For time invariant transition probabilities, $p(i,t)=p(i)$ and for
fixed $N$, we can write
\begin{equation}
P(t) = P(0) M^t.
\label{dynamics}
\end{equation}
Our formalism is a simplified variation of previous statistical
approaches, which use renewal models for neuronal
firing~\cite{gerstner1992,gerstner1995,meyer2002}.  In those
approaches, the probability for a given neuron to fire depends on the
refractory dynamics and inputs received over the time interval since
the last time that particular neuron fired.   Our model assumes that
the  neurons are completely indistinguishable so the only relevant
measure of the network dynamics is the number of neurons active within
a temporal epoch.  Thus the probability of a neuron to fire depends
only on the network activity in a previous temporal epoch.  This
simplifying assumption allows us to compute the PDF
of the network activity, the first two moments, and the
autocorrelation function analytically.

%Our approach differs from mean-field theory because it does not insist on self-consistency between the activity of an individual neuron and the surrounding network.  Instead, it presumes that the neuron is only sensitive to some aspect of network activity, which in our case is the total number of active neurons, and then calculates the neuronal response completely.  Hence, as will be shown in the ensuing,  the firing properties of an individual neuron need not reflect the network statistics at all.

\section{Time independent Markov model}

We consider a network of size $N$ with a time independent response
function, $p(i,t)=p(i)$, so that (\ref{dynamics}) specifies the 
temporal evolution of the PDF of the network activity.  
If $0 < p(i) < 1$ for all $i \in [0,N]$ (i.e. the probability of
firing is never zero nor one), then the MTM is a positive matrix (i.e. it
has only non-zero 
entries). The sum over a row of the MTM is unity by construction.
Thus the maximum row sum matrix norm of the MTM is unity implying that the
spectral radius 
is also unity.  Hence,  by the Frobenius-Perron Theorem, the maximum
eigenvalue is unity and it is unique.   This implies that for all
initial states of the PDF, $P(t)$ converges to a unique left eigenvector $\mu$
called the invariant measure, which satisfies the relation 
\begin{equation}
\mu = \mu M.
\label{eq:invariant}
\end{equation}
The invariant measure is the attractor of the network dynamics and
the equilibrium PDF.  Given that the spectral radius of $M$ is unity, convergence to the equilibrium will be exponential at a rate governed by the penultimate eigenvalue.  We note
that if the MTM is not positive then there may not be an invariant measure.
For example if the probability of just one possible transition is zero, then
the PDF may never settle to an equilibrium.  

The PDF specifies {\em all} the information of the system and can be
found by solving (\ref{eq:invariant}).  From (\ref{dynamics}), we see that the
invariant measure is given by the column average over any arbitrary
PDF of $\lim_{t\rightarrow \infty} M^t$.  Hence, the infinite
power of the MTM must be a matrix with equal rows, each row being the
invariant measure.  Thus, a way
to approximate the invariant measure to arbitrary precision is to take one row of a large
power of the MTM.  The higher the power, the more precise the
approximation.  Since the convergence to equilibrium is exponential, a very accurate 
approximation can be obtained easily.

\subsection{Mean and variance}
In general, of
most interest are 
the first two moments of the PDF so we derive expressions for
the expectation value and variance of any function of the network
activity. Let $X\in [0,...N]$ be a random variable representing the network
activity and $f$ be a real valued function of $X$.  The expectation
value and variance of $f$ at time $t$ is thus 
\begin{equation}
\langle f(X) \rangle_{t} = \sum_{k=0}^N f(k) P_k(t)
\label{eq:expect}
\end{equation}
and
\begin{equation}
Var_{t}(f(X)) = \langle f(X)^2 \rangle _{t} - \langle f(X) \rangle
_{t}^2.
\label{eq:var}
\end{equation}
where we denote the $k$th element of vector $P(t)$ by
$P_k(t)$. 
We note that in numerical simulations, we will implicitly assume ergodicity at equilibrium. That is, we will consider that the expectation value over all the possible outcomes to be equal to the expectation over time.  

Inserting (\ref{M}) into (\ref{eq:dyn1}) and using the definition of
the mean of a binomial distribution, the mean activity
$\langle X \rangle_t$ has the form
\begin{equation}
\label{eq:markov_mean}
\langle X \rangle_{t}= N \langle p(X) \rangle_{t-1}.
\end{equation}
The mean firing rate of the network is $\langle X \rangle_{t}/\Delta t$.
We can also show that the variance is given by
\begin{equation}
\label{eq:markov_variance}
Var_t(X) = N\langle p(1-p) \rangle _{t-1} + N^2Var_{t-1}(p).
\end{equation}
Details of the calculations for the mean and variance are in
Appendix \ref{app:mean}.   Thus the mean and variance of the network activity are expressible in terms of the mean and variance of the response function.    At equilibrium,  (\ref{eq:markov_mean}) and (\ref{eq:markov_variance}) are
$\langle X \rangle _\mu = N\langle p \rangle _\mu$  and 
$Var_\mu(X) = N\langle p(1-p) \rangle _\mu + N^2Var_\mu(p)$,
respectively.  Given an MTM, we
can calculate the invariant measure and then from that obtain all the
moments.

If we suppose the response function to be linear, we can compute the mean and variance in closed-form.  Consider the linear response function
\begin{equation}
p(X) = p_0 + \frac{(q-p_0)}{Nq}X 
\label{eq:linear}
\end{equation}
for $X\in [0,N]$.
Here $p_0 \in [0,1]$ is the probability of firing  for no inputs and  $(q-p_0)/Nq$ is the slope of the response function.  Inserting (\ref{eq:linear}) into (\ref{eq:markov_mean}),
gives  
\begin{equation}
\langle X \rangle _\mu = Np_0 +\frac{(q-p_0)}{q}\langle X \rangle _\mu.
\end{equation}
Solving for $\langle X \rangle_\mu$ gives the mean activity
\begin{equation}
\langle  X \rangle _\mu = Nq,
\label{eq:linmean}
\end{equation}
%which satisfies (\ref{eq:fp}).
Substituting (\ref{eq:linear}) into (\ref{eq:markov_variance}) leads to
the variance 
\begin{equation}
Var_\mu(X) =  N\frac{q(1-q)}{1 - \lambda^2 + \lambda^2/N }
\label{eq:linvar}
\end{equation}
where $\lambda=(q-p_0)/q$.
The details of these calculations are given in Appendix \ref{app:mean}.

The expressions for the mean and variance give several insights into the model.  From (\ref{eq:linear}), we see that the mean activity (\ref{eq:linmean}) satisfies the condition 
\begin{equation}
\label{eq:fp}
p(Nq)=q.
\end{equation}
Hence, on a graph of the response function versus the input,  the mean response
probability is given by the intersection of the response
function and a line of slope $1/N$, which we denote the {\em diagonal}.   Using (\ref{eq:linmean}),  (\ref{eq:fp}) can be re-expressed as
\begin{equation}
\langle X \rangle_\mu = N p(\langle X \rangle_\mu)
%\label{eq:fp}
\end{equation}
In equilibrium, the mean response of a neuron to $\langle X
\rangle$ neurons firing is $\langle X \rangle/N$. Hence, for a linear
response function, the mean network activity obeys the
self-consistency condition of mean-field theory.  This can be
expected because of the assumption 
of statistical homogeneity.

The variance (\ref{eq:linvar}) does not match a mean-field
theory that assumes uncorrelated statistically independent firing.
In equilibrium, the mean probability to fire of a single neuron
in the network is  $q$. Thus, each neuron
obeys a binomial distribution with variance $q(1-q)$. 
Mean-field theory would predict that the network variance would then be
given by $Nq(1-q)$.
Equation (\ref{eq:linvar}) shows that the variance exceeds
the statistically independent result by a factor that is size
dependent, except 
for $q=p_0$, which corresponds to an unconnected network.  
Hence, the variance of the network activity cannot be discerned from
the firing characteristics of a single neuron.  The network could
possess very large correlations while each constituent neuron would exhibit
uncorrelated firing.

When $N>>\lambda^2$, the variance scales as $N$ but for small $N$, there
is a size-dependent correction. The coefficient of
variation approaches zero as $1/\sqrt{N}$ as $N$ approaches infinity.
When $\lambda=1$, the slope of the response function is $1/N$ and
coincides with the diagonal. This is a critical point where the
equilibrium state becomes a line attractor~\cite{seung1996}. In the
limit of $N\rightarrow\infty$, the variance diverges at the critical
point.  At criticality, the variance of our model has the form
$N^2 q(1-q)$, which diverges as the square of $N$.  
Away from the critical point, in the limit as $N$ goes to infinity,
the variance has the form $(Nq(1-q))/(1-\lambda^2)$.  Thus, the
deviation from mean-field theory is present for all network sizes and
becomes more pronounced near criticality.

The mean-field solution of (\ref{eq:fp}) is not a strict fixed point of the
dynamics (\ref{dynamics}) per se.  For example, if the response
function is near criticality, possesses discontinuities, or crosses
the diagonal multiple times then the solution of (\ref{eq:fp}) may not
give the correct mean activity.   
Additionally, if the slope of the
crossing has a magnitude greater than $1/N$, then the crossing point would be
``locally unstable''.  Consider, small perturbations of the mean around
the fixed point given by (\ref{eq:linmean}) 
\begin{equation}
\langle X \rangle_t = N q + v(t)
\label{fp2}
\end{equation}
The response function (\ref{eq:linear}) then takes the form
\begin{equation}
p(X)= q + \frac{\lambda}{N} v
\label{lin2}
\end{equation}
where $\lambda=(q-p_0)/q$.  Substituting (\ref{fp2}) and (\ref{lin2}) into (\ref{eq:markov_mean}) gives
\begin{equation}
\langle v \rangle_t= \lambda \langle v \rangle_{t-1}
\end{equation}
Hence, $v$ will approach zero (i.e. fixed point is stable) only if $|\lambda|<1$.  

Finally, we note that
in the case that there is only one solution to (\ref{eq:fp})
(i.e. response function crosses the diagonal only once), if the
function is continuous and monotonic then it can only cross with
$\lambda<1$ (slope less than $1/N$) since $0<p<1$.  Thus, a continuous
monotonic increasing response function that crosses the diagonal only once, has a stable mean activity given by this crossing.  This mean activity corresponds to the mode of the invariant measure.

%By definition, $0<p< 1$  so $Np$ lies in $(0,N)$. Theoretically, the response function can have any shape.  In practice, it will be monotonic since it describes the probability of firing of one neuron according to an increase of input. Hence, this function will be decreasing for an inhibitory network and increasing for an excitatory one. A constant response function describes an uncoupled network.  

We can estimate the mean and variance of the activity for a smooth nonlinear response function that crosses the diagonal once by linearizing the response function around the crossing point ($q=p(Nq)$).  Hence, near the intersection
\begin{equation}
p(X) = q + \lambda \frac{(X-Nq)}{N}.
\end{equation}
where  $\lambda = N\frac{\partial p}{\partial X}(Nq)$.  Using (\ref{eq:linmean}) and (\ref{eq:linvar}) then gives
\begin{eqnarray}
\label{eq:finalmean}
\langle X \rangle _\mu &=& Nq \\
Var_\mu(X) &=& \frac{Nq(1-q)}{1 - \lambda^2 + \lambda^2/N }
\label{eq:finalvariance}
\end{eqnarray}
Additionally, the crossing point is stable if $|\lambda|<1$.  These linearized estimates are likely to break down near criticality (i.e. $\lambda$ near one).

\begin{figure}[htbp]
\begin{center}
\includegraphics[width=4.25in]{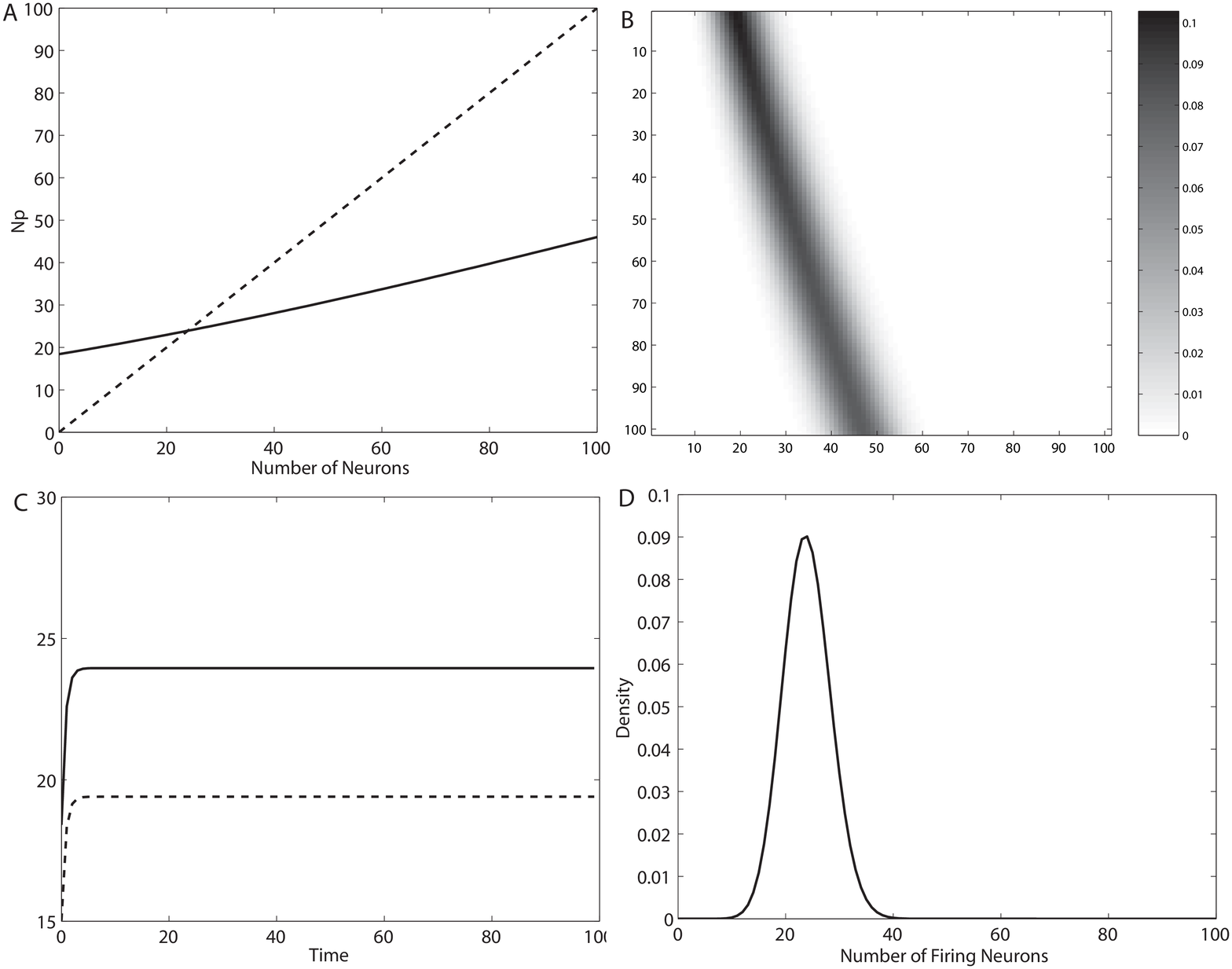}
\end{center}
\caption{Example for response function $p(n) =
  \frac{1}{2\pi}\int_{\frac{\theta - I -
      Jn}{\sigma}}^{\infty}e^{-\frac{x^2}{2}}dx$ with $N=100$, $\theta
  =1$, $I=0.1$, $J=1.5/N$ and $\sigma=1.0$. A) $Np(n)$ (solid line)
  and the diagonal (dashed line). B) The associated Markov Transition
  Matrix (in gray levels). C) The evolution of the mean network
  activity  (solid line) and the variance (dashed line). The
  steady-state is quickly reached and the activity corresponds to the
  crossing point between $Np(n)$ and the diagonal. D) The invariant
  measure (i.e the PDF of the network activity). }
\label{fig:example}
\end{figure}

We show an example of a response function and resulting MTM in 
Figure \ref{fig:example}. Starting with a
trivial density function (all neurons are silent), we show the time
evolution of the mean
and the variance in figure \ref{fig:example}C.   We show the invariant
measure in figure \ref{fig:example}D.  The equilibrium state is given by the
intersection of the response function with the diagonal. We see that the mode of the invariant measure is aligned with the mean activity given by the crossing point.

\subsection{Autocovariance and autocorrelation functions}
We can also compute the autocovariance function of the network activity:
\begin{equation}
\rm{Cov}(\tau )= \langle (X(t)-\langle X\rangle_\mu ) (X(t+\tau)-\langle X\rangle_\mu )\rangle=\langle X(t)X(t+\tau)\rangle-\langle X^2\rangle_\mu.
\end{equation}
Noting that
\begin{equation}
\langle X(t)X(t+\tau)\rangle=\sum_j\sum_k j k p(X(t+\tau)=k | X(t)=j)p(X(t)=j)
\end{equation}
where $p(X(t+\tau)=k | X(t)=j)=M_{jk}^\tau$,
we show in Appendix \ref{app:ac}, that at equilibrium
\begin{equation}
\rm{Cov}(\tau )= \lambda^\tau Var_\mu(X)
\label{eq:autoone}
\end{equation}
where $\lambda$ is the {\em slope factor} ($N$ times the slope) of the
response function evaluated at the crossing point with the diagonal.
The autocorrelation function $AC(\tau)$ is simply $\lambda^\tau$. 

The correlation time is given by $1/\ln \lambda$.  At the critical
point ($\lambda=1$), the correlation time becomes infinite.  
For a totally
decorrelated network, $\lambda=0$, giving a Kronecker delta function for
the autocorrelation  as expected.
For an inhibitory
network ($\lambda<0$), the autocorrelation exhibits 
decaying oscillations with period equal to the time step of the Markov
approximation (i.e. $\rm{Cov}(1)<0$).  (Since we have assumed $p>0$,
an inhibitory network in this formalism is presumed to be driven by an external current
and the probability of firing decreases when other neurons in the
local network fire.)
  We stress that these correlations are only
apparent at the network level.  Unless the rate is very high, a single
neuron will have a very short correlation time.

%We now consider the speed of convergence. If we use a Jordan decomposition on the orthogonal space of the eigenvector, we find that the speed of convergence toward the invariant measure is a product of a polynomial and an exponential. The rate constant of the exponential is the largest magnitude eigenvalue different from one and the degree of this polynomial is the multiplicity of this eigenvalue minus one . In the case of only one equilibrium, the computation of the autocorrelation function suggests that the rate is also the slope of the response function: the slope is the second largest eigenvalue (in absolute value) of the Markov transition matrix. 

\subsection{Multiple Crossing Points}

\begin{figure}[htbp]
\begin{center}
\includegraphics[width=4.25in]{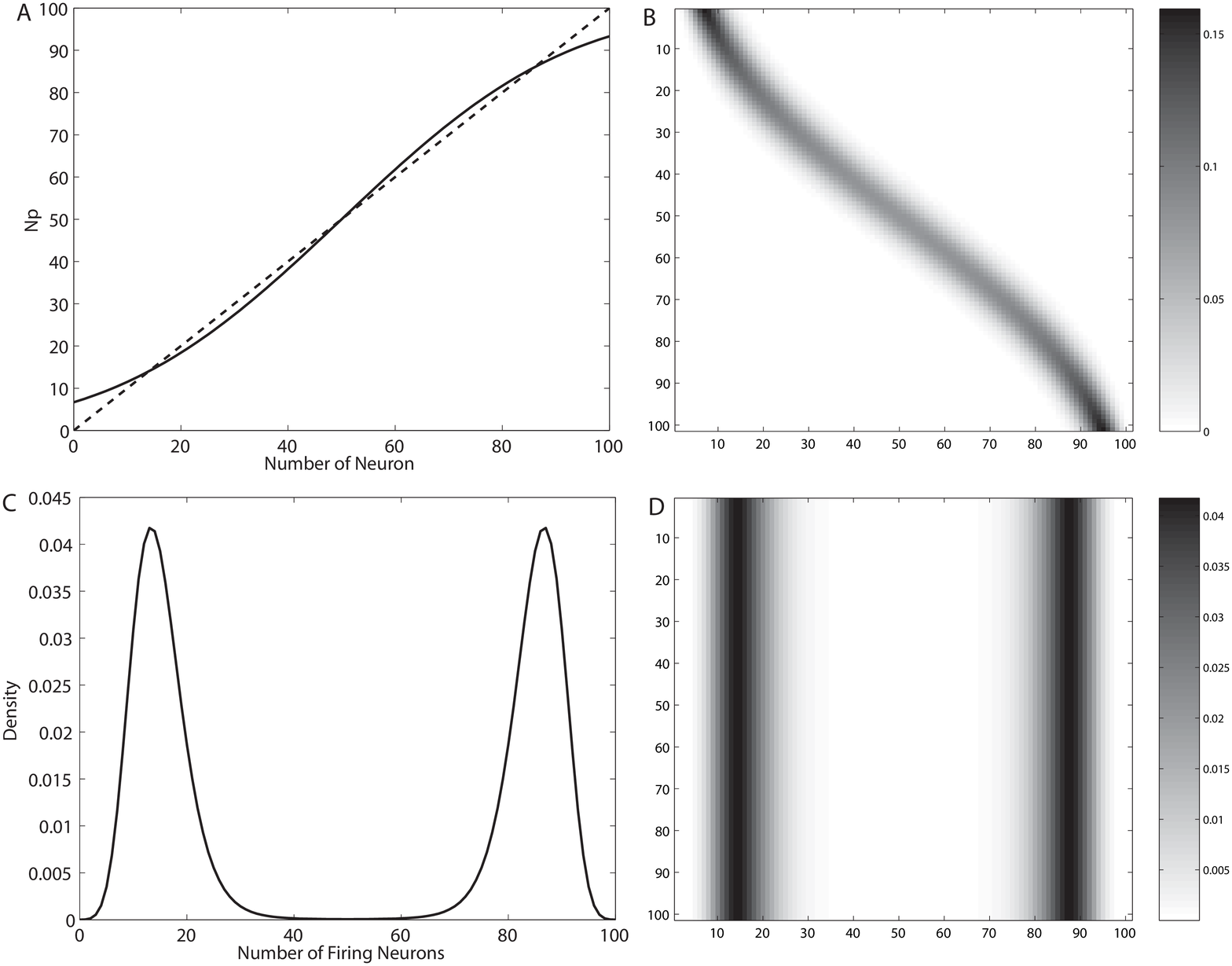}
\end{center}
\caption{Example where the response function crosses the diagonal
  three times. A) $Np(n)$ (solid line) and diagonal (dashed line).  B)
  The associated MTM. C) The invariant measure. D) MTM$^{10^5}$. All
  rows are equal to the invariant measure. Parameters are $N=100$, $I=0.1$, $J=1.8/N$ and $\sigma=0.6$. }
\label{fig:example2}
\end{figure}

For dynamics where $0<p<1$,  if the response function is continuous
and crosses the diagonal more than once,  then the number of crossings
must be odd.  Consider the case with three crossings with $p(0)<p(N)$.
If the slopes of all the crossings are positive (i.e. a sigmoidal
response function), then the slope factor $\lambda$ of the middle
crossing will be 
greater than one while the other two crossings will
have $\lambda$ less than  one implying two stable crossing points and
one unstable one in the middle. 
Figure~\ref{fig:example2}A and B provide an example of such a response function and its associated MTM. The invariant measure is shown in figure \ref{fig:example2}C.     We see
that it is bimodal with local peaks at the two crossing points with
$\lambda$ less than one.  The invariant measure is well approximated
by any row of a large power of the MTM (figure \ref{fig:example2}D).
If the stable crossing points are separated
enough (beyond a few standard deviations), then we can estimate the
network mean and variance by combining the two local mean and variances. 
Assuming that the stable crossing points are $q_1$ and $q_2$ ($q_1<
q_2$) with slope factors  $\lambda_1$ and $\lambda_2$.   Then the mean and
the variance are: 
\begin{eqnarray}
\label{eq:finalmean2}
\langle X \rangle _\mu &=& N\frac{q_1+q_2}{2}\\
Var_\mu (X) &=& \frac{N}{2}\left(\frac{q_1(1-q_1)}{1 - \lambda_1^2 + \lambda_1^2/N } + \frac{q_2(1-q_2)}{1 - \lambda_2^2 + \lambda_2^2/N }\right) + N^2\left(\frac{q_1-q_2}{2}\right)^2
\label{eq:finalvariance2}
\end{eqnarray}
%Our model allows us
%to estimate the firing statistics according to the size of the
%network as long as the response function can be derived or computed.
%In the following, we show example applications of the Markov
%formalism to various neural models.

\section{Fast leak spiking model}

We now consider a simple neuron model where the
firing state obeys
\begin{equation}
V_i(t) = \rm{H}(I(t) + \frac{1}{N}\sum_{k=1}^{N}J_{ik}V_k(t-1) + S(t)-\theta)
\label{eq:stateequation}
\end{equation}
where $\rm{H}(\cdot)$ is the Heaviside step function, $I(t)$ is an input
current, $S(t)=N(0,\sigma)$ is uncorrelated Gaussian noise,
$\theta$ is a threshold, and $J_{ik}$ is the connection strength
between neurons in the network.  When the combined input of the neuron
is below threshold, $V_i=0$ (neuron is silent) and when the input exceeds
threshold, $V_i=1$ (neuron is active). This is a simple version of a spike response or renewal model~\cite{gerstner1992,gerstner1995,meyer2002}.

In order to apply our Markov formalism, we need to construct the
response function.   We assume
first that the connection matrix has constant entries $J$.  We will also 
consider disordered connectivity  later. The response function is the
probability that a neuron will fire given that $n$ neurons fired
previously.  The
response function at $t$ for the neuron is then given by 
\begin{equation}
\label{eq:pfast}
p(n,t) = \frac{1}{\sqrt{2\pi}} \int_{\frac{\theta - I(t) - nJ/N}{\sigma}}^{\infty} e^{-x^2/2}dx 
\end{equation}
If we assume a time independent input then we can construct the MTM using
(\ref{M}).  The equilibrium PDF is given by (\ref{eq:invariant}), and
can be approximated by one row of a large power of the  MTM.
\subsection{Mean, variance and autocorrelation}

Imposing the self-consistency condition (\ref{eq:fp}) on the mean activity $Nq$, where $q=\langle p \rangle_\mu$, and $p$ is given by (\ref{eq:pfast}) leads to 
\begin{equation}
\label{eq:self_m}
q=\frac{1}{\sqrt{2\pi}} \int_{\frac{\theta - I - qJ}{\sigma}}^{\infty} e^{-x^2/2}dx.
\end{equation}
Taking the derivative of $p(n,t)$ in (\ref{eq:pfast}) at $n=Nq$ gives
\begin{equation}
\label{eq:self_J}
\lambda = \frac{J}{\sigma\sqrt{2\pi}}e^{-\frac{{(\theta-I-Jq)}^2}{2\sigma^2}}.
\end{equation} 
Using this in  (\ref{eq:finalvariance}) gives the variance.

Equation (\ref{eq:self_J}) shows the  influence of noise and recurrent
excitation on the slope factor and hence variance of the network activity.  For $\lambda<1$, the variance (\ref{eq:finalvariance}) increases with $\lambda$.
An increase in excitation $J$ increases $\lambda$ for large $J$ but decreases $\lambda$ for small $J$.  Similarly, a decrease in noise $\sigma$ increases $\lambda$ for large $\sigma$ but decreases it for small $\sigma$.  Thus, variability in the system can increase or decrease with  synaptic excitation and external noise depending on the regime.  The autocovariance function, from  (\ref{eq:autoone}) is 
\begin{equation}
Cov(\tau) = \lambda^\tau \frac{Nq(1-q)}{1-\lambda^2 + \lambda^2/N }.
\label{cov}
\end{equation} 
\subsection{Bifurcation diagram}
Bifurcations occur when the stability of the mean-field solutions (i.e. crossing points) for the mean activity  changes.
We can construct the bifurcation diagram on the $I-J$ plane by locating crossing points where $\lambda=1$.  We first consider rescaled parameters
$\mathcal{I} = (\theta - I)/\sigma$ and $\mathcal{J}=J/\sigma$.  Then
(\ref{eq:self_m}) becomes
\begin{equation}
\label{eq:self_scale}
q = \frac{1}{\sqrt{2\pi}} \int_{\mathcal{I} - q\mathcal{J}}^{\infty} e^{-x^2/2}dx 
\end{equation}
and (\ref{eq:self_J}) becomes  
\begin{equation}
\label{eq:lambda_scale}
\lambda = \frac{\mathcal{J}}{\sqrt{2\pi}}e^{-\frac{{(\mathcal{I}-\mathcal{J}q)}^2}{2}}
\end{equation} 
Solving (\ref{eq:lambda_scale}) for $q$ and 
inserting into
(\ref{eq:self_scale}) gives
 
\begin{equation}
\mathcal{I} = \frac{\mathcal{J}}{\sqrt{2\pi}} \int_{\pm \sqrt{\ln(\frac{\mathcal{J}^2}{2\pi\lambda^2})}}^{\infty} e^{-x^2/2}dx \pm \sqrt{\ln(\frac{\mathcal{J}^2}{2\pi\lambda^2})} 
\label{eq:bifequ}
\end{equation}

\begin{figure}[htbp]
\begin{center}
\includegraphics[width=4.25in]{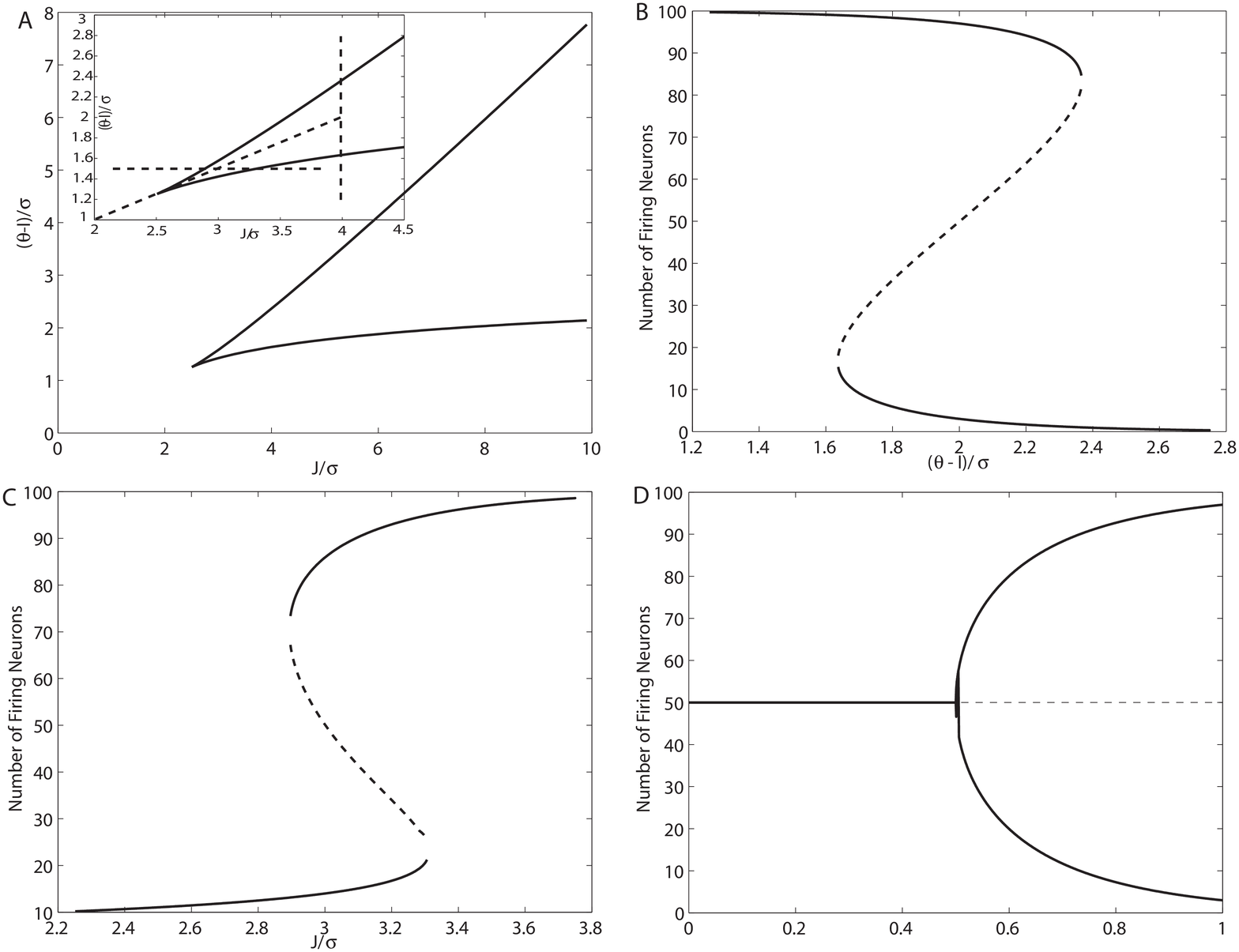}
\end{center}
\caption{A) The two dimensional bifurcation diagram ($\lambda=1$) as a function of $\frac{\theta-I}{\sigma}$ and $\frac{J}{\sigma}$. Inset: the traversal lines of the various one dimensional bifurcation diagrams shown in B) vertical, C) horizontal and D) oblique.  }
\label{fig:bif}
\end{figure}

The bifurcation points are obtained by setting  $\lambda=1$
in (\ref{eq:bifequ}) and evaluating the integral.  The solutions have two branches that satisfy the
conditions
$\mathcal{J}\geq \sqrt{2\pi}$ and $\mathcal{I} \geq \sqrt{\pi/2}$.
Figure \ref{fig:bif}A shows the two dimensional bifurcation diagram
on the $\mathcal{I}-\mathcal{J}$ plane.  The intersection of
the two branches at $\mathcal{J}= \sqrt{2\pi}$ and $\mathcal{I}
=\sqrt{\frac{\pi}{2}}$ is a co-dimension two {\em cusp} bifurcation
point (i.e. satisfies $Np'(x)=1$, $Np''(x)=0$ with genericity
conditions \cite{kuznetsov1998}).  Between the two branches 
 there are three crossing points (two of which
are stable), outside there is one stable crossing point. 
Each traversal of a branch yields a saddle node bifurcation as seen in figures \ref{fig:bif}B and C for the vertical and horizontal traversals (shown in the inset to figure \ref{fig:bif}A) respectively.  Traversing through the critical point along the diagonal line in the inset of figure \ref{fig:bif}A shows a pitchfork bifurcation.  
We have assumed that $J>0$ (i.e excitatory network).  By taking
$J\rightarrow -J$ we obtain the same bifurcation diagram for the inhibitory
case and the above conditions still hold using the absolute value of $J$. 

\begin{figure}[htbp]
\begin{center}
\includegraphics[width=4.25in]{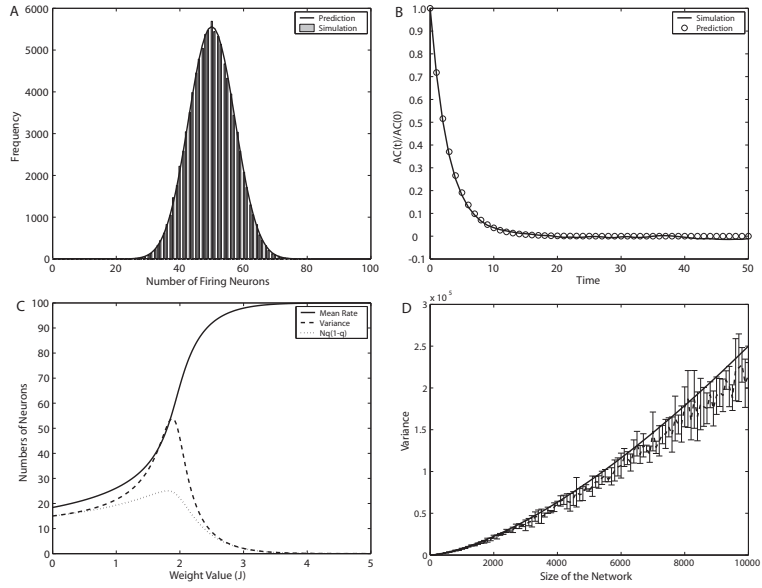}
\end{center}
\caption{A) Numerical and theoretical PDF of the network activity at equilibrium for $N=100$, $\theta=1$, $I=0.1$, $\sigma=0.8$ and $J=1.8$. The theoretical PDF was obtained by taking one row of $MTM^{100}$. B) Numerical and theoretical autocorrelation function for same parameters as A). Solid line is $\lambda^t$ with the predicted $\lambda$ from equation (\ref{eq:self_J}). C) Dependence of the mean (solid line) and variance (dashed line) with $J$.  The mean-field variance is $Nq(1-q)$ (dotted line). D) Evolution of the variance when $\lambda=1$ versus $N$ for $I=1-\sqrt{2\pi}$, $J=\sqrt{2\pi}$, $\sigma=1$, $\theta=1$.  The solid line is $N^{3/2}q(1-q)$.}
\label{fig:anares}
\end{figure}

We compare our analytical estimates from the Markov model with numerical simulations of the fast leak model  (\ref{eq:stateequation}) in figure~\ref{fig:anares}.
Figure \ref{fig:anares}A shows that the equilibrium PDF derived from
the MTM matches the PDF generated from the simulation. 
Figure \ref{fig:anares}B shows a comparison of $\lambda^\tau$ versus
the simulation autocorrelation function.   
Figure \ref{fig:anares}C shows the mean and variance of the activity
versus the connection weight $J$. The theoretical results match the
simulation results very well.   On the same plot, we show the mean-field theory estimate where the neurons are assumed to be completely uncorrelated.  For low recurrent excitation and hence low activity, the correlations can be neglected
and the system behaves like an uncorrelated neural network.  However, as
the excitation and firing activity increase,  correlations
become more significant and the variance grows quicker than the uncorrelated result.  For very strong connections, the firing activity
saturates and the variance drops.  In that case, mean-field theory is again an
adequate approximation.
%When $\lambda=1$,  we find that the variance grows as $N^2$ while the mean-field theory predicts it grows as $N$.   
Figure \ref{fig:anares}D shows the variance versus $N$ for a network at the critical point, $\mathcal{J}= \sqrt{2\pi}$ and $\mathcal{I}=\sqrt{\frac{\pi}{2}}$ where $\lambda=1$ at the crossing point.  We see that the variance grows approximately as $N^{3/2}$. This is faster than the mean-field result but slower than the prediction for a linear response function.  The breakdown of the linear estimate is expected near criticality. 

\begin{figure}[htbp]
\begin{center}
\includegraphics[width=4.25in]{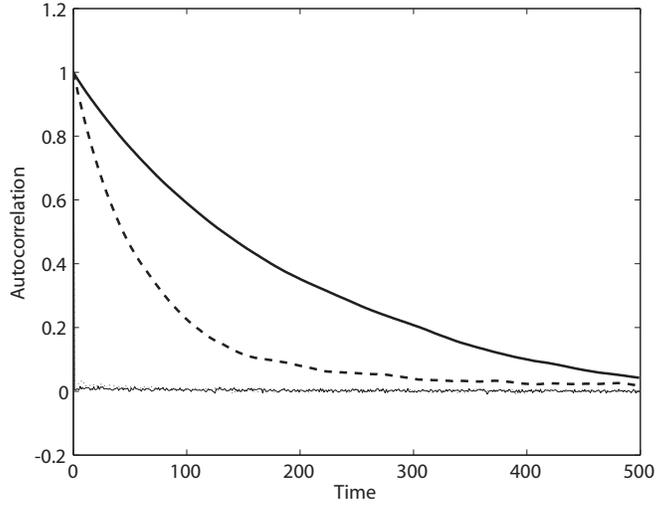}
\end{center}
\caption{Autocorrelation of the network and one neuron at a critical
  point ($\lambda=1$). The solid line is for a network of 10,000
  neurons and the dashed line is for a network of 1000 neurons. 
  Crosses and dots are the autocorrelation of one neuron for $N=1000$
  and $N=10,000$ respectively. The
  autocorrelation for one neuron decays almost immediately.  Parameters are $\sigma=1$, $\theta=1$,
  $J=\sqrt{2\pi}$ and $I=1-\sqrt{2/\pi}$.} 
\label{fig:auto_crit}
\end{figure}

Figure \ref{fig:auto_crit} shows the simulation autocorrelation for
both the network and one neuron (chosen randomly) for  $N=1000$ and
$N=10,000$. The parameters were chosen so the network was at a
critical point ($\lambda=1$).  We see that the correlation time of an
individual neuron is near zero and much shorter than that of the
network.   With increasing $N$, we find that the correlation time {\em increases} for the network and {\em decreases} for the single neuron.  Thus, a single neuron can be completely uncorrelated while the network has very long correlation times.

\subsection{Bistable regime}

\begin{figure}[htbp]
\begin{center}
\includegraphics[width=4.25in]{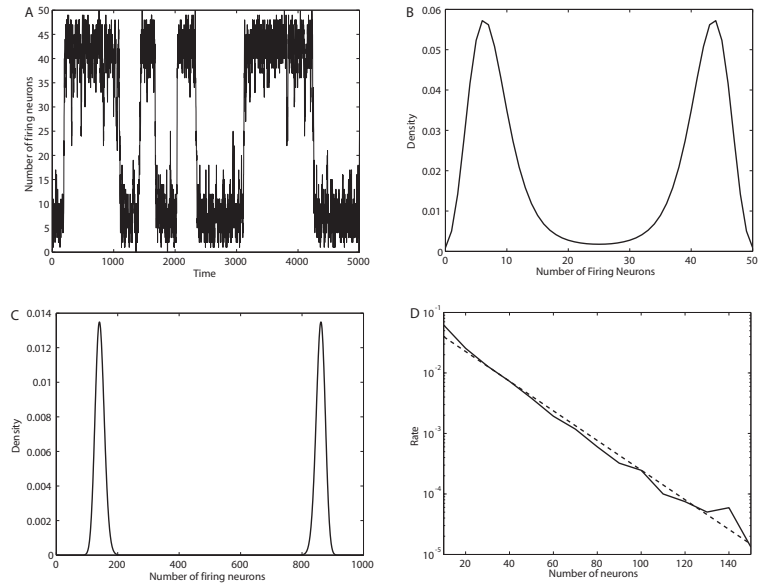}
\end{center}
\caption{A) The firing activity over time of a bistable network of 50 neurons.  The switching between the two states (high and low activity) is spontaneous. B) Theoretical PDF for $N=50$.  C) Theoretical PDF for $N=1000$. D) Switching rate depends exponentially on  $N$ (solid curve).  Dashed line is a linear fit on semi-log scale.
Parameters are: $J=1.8$, $\sigma=0.6$, $I=0.1$, $\theta=1$.}
\label{fig:bimodal}
\end{figure}

In the regime where there are three crossing points, the 
invariant measure is bimodal.  Hence, the network activity is effectively bistable.
Bistable network activity has been proposed as a model of
working memory \cite{compte2000,laing2001,gutkin2001,compte2006}. The 
two local equilibria are locally stable but due to the finite-size
fluctuations,  there can be spontaneous
transitions from one 
state to the other.
Figure \ref{fig:bimodal}A shows the
activity over time for a bistable network with $N=50$.  The
associated PDF is shown in figure \ref{fig:bimodal}B.  
No external input has been added to the network.
For $N=1000$, the switching rate is sufficiently low to be
neglected. The resulting probability density function for this network
is displayed on figure \ref{fig:bimodal}C. 

We can estimate the scaling behavior on $N$ of the state switching
rate and hence lifetime of a memory by supposing the finite-size induced
fluctuations can be approximated by uncorrelated Gaussian noise.  We
can then map the switching dynamics to barrier hopping of a stochastically forced particle
in a double 
potential well $V(x)$ with a stationary PDF given by the invariant measure
$\mu$.   The switching rate is given by the formula
\cite{risken1989} 
\begin{equation}
\label{eq:jump_rate}
r = (2\pi)^{-1}\sqrt{V''(c)|V''(a)|}e^{-\frac{(V(a)-V(c))}{\sigma^2}}
\end{equation}  
where $V=-\sigma^2\ln(\mu)$, $\mu$ is the
invariant measure, $c$ is a stable crossing point,
$a$ is the  unstable crossing point, and
$\sigma^2$ is the forcing noise variance around $c$, which we take to be proportional to the variance of the network activity.
For the general case, we cannot derive this rate
analytically but we can obtain an estimate by assuming that the
invariant measure is a sum of two Gaussian PDFs
whose means are exactly the two stable crossing points.
\begin{equation}
\mu(x) = \frac{\exp\left(-\frac{(x-Nq_1)^2}{2\sigma^2_1}\right)}{2\sqrt{2\pi}\sigma_1}+\frac{\exp\left(-\frac{(x-Nq_2)^2}{2\sigma^2_2}\right)}{2\sqrt{2\pi}\sigma_2},
\end{equation} 
where  $\sigma^2_i = \frac{Nq_i(1-q_i)}{1-\lambda^2_i+\lambda^2_i/N}$ for $i=1,2$.
Using this in (\ref{eq:jump_rate}) then gives
\begin{equation}
r \propto \Omega e^{-kN},
\label{eq:rate_approx}
\end{equation}
for a constant $k$.
The switching rate is an exponentially decreasing function of $N$. 
We computed the switching rate  for $N \in [0;150]$  for the parameters of
the bistable network above. The results are displayed on figure  
\ref{fig:bimodal}D. For the parameters in \ref{fig:bimodal}D, a fit
shows that $k$ is of order 0.05.  For these parameters, switching
between states will not be important when $N>>1/k\sim 20$.  

\subsection{Disordered connectivity}
\begin{figure}[htbp]
\begin{center}
\includegraphics[width=4.25in]{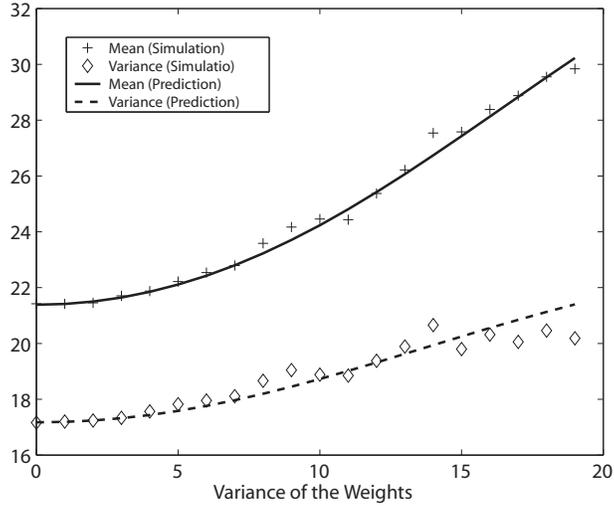}
\end{center}
\caption{Mean (crosses) and variance (diamonds) from a numerical simulation of the fast leak model for $N=100$ neurons as a function of the variance of the random disordered connection weights.  Theoretical mean (solid line) and variance (dashed line) match well with numerical simulation values.  Parameters are $J=0.5/N$, $I=0.1$, $\sigma=1.0$, $\theta=1.0$.}
\label{fig:random}
\end{figure}

We now consider the dynamics for disordered (random) connection
weights $J_{ij}$ drawn from a Gaussian distribution
($N(\bar{J},\sigma^2_J)$).   Soula et al.~(2006) showed that the input
to a given neuron for a disordered network can be approximated by
stochastic  input.  We thus choose random input
drawn from the distribution,
$N(k\bar{J},k\sigma^2_J)$, for $k$ neurons having fired in the previous
epoch. The response function (for time independent
input) then obeys
\begin{equation}
p(n) = \frac{1}{\sqrt{2\pi}} \int_{\frac{\theta - I - n\bar{J}/N}{\sqrt{\sigma^2+n\sigma^2_J}/N}}^{\infty} e^{-x^2/2}dx.
\end{equation}
Applying the self-consistency condition (\ref{eq:fp}) for the mean firing probability $q$  gives
\begin{equation}
q= \frac{1}{\sqrt{2\pi}} \int_{\frac{\theta - I - q\bar{J}}{\sqrt{\sigma^2+q\sigma^2_J}}}^{\infty} e^{-x^2/2}dx 
\end{equation}
The variance is given by (\ref{eq:finalvariance}) with 
\begin{equation}
\lambda =\frac{1}{\sqrt{2\pi}} \left(\frac{\bar{J}}{\sqrt{\sigma^2+q\sigma^2_J}} - \sigma^2_J\frac{\theta - I - q\bar{J}}{2(\sigma^2+q\sigma^2_J )^{3/2}}\right) e^{-\frac{{(\theta-I-\bar{J}q)}^2}{2(\sigma^2+q\sigma^2_J )}}
\end{equation}
Figure \ref{fig:random}  shows the activity mean and variance as a
function of the connection weight variance.  There is a close match
between the prediction and the results generated from a direct
numerical simulation with 100 neurons.  We note that a possible
consequence of disorder in a neural network is a spin glass where many
competing activity states could co-exist and so the resulting activity
is strongly dependent on the initial
condition~\cite{hopfield1982}.  However, we believe that the stochastic
forcing inherent in our network is large enough to overcome any spin
glass effects.  In the language of statistical mechanics, our system
is at a high enough temperature to smooth the free energy function.

%In figures \ref{fig:random} (bottom), we examine the effect of weight
%heterogeneity on a bistable network. On the left, the modes are
%computed versus the variance of the weights and on the right the
%associated slopes factors. As shown, the network remains in bistable
%conditions (two modes) for a wide range of weight
%heterogeneity. There is therefore, no need for a fine tuning to
%obtain a network in this regime.  

\section{Integrate-and-fire model}

We now apply our formalism
to a more biophysical integrate-and-fire model.
We consider two versions. The first
treats synaptic input as a current source and the second
considers synaptic input in the form of a conductance change.  We apply our Markov model with a discrete time step chosen to be the larger of the membrane or synaptic time constants.

\subsection{Current-based synapse}
We first consider an integrate-and-fire model where the 
synaptic input is applied as a current.  The membrane potential obeys
\begin{equation}
\tau\frac{Dev_i}{dt}= I - V + \frac{J}{N}\sum_{t^f<t}^{N}\alpha(t-t^f) + Z(t)
\end{equation}
where $\alpha(t)=e^{-\frac{t}{\tau_s}}/\tau_s$ if $t>0$ and zero
otherwise, $I$ is a constant input, $Z$ is an uncorrelated zero-mean white
noise with variance $\sigma$, $J$ is the synaptic strength coefficient, $N$ is the network size, and  $t^f$ are the  firing times of all neurons in the network.
These firing times are computed whenever the membrane
potential crosses a threshold $V_\theta$, whereupon the potential
is reset to  $V_{Rs}$. After firing, neurons are prevented from firing
during an absolute refractory period $r$.
The parameter values are
$\tau = 1$ ms, $\tau_s = 1$ ms,$V_r = -65$ mV, $V_\theta = -60$ mV,
$V_{Rs}=-80$ mV and $r=1$ ms. 

 The network dynamics of this model had been
studied previously in detail with a Fokker-Planck 
formalism~\cite{brunel2000a,brunel2006}.  These analyses 
 showed that
if the input current is an uncorrelated white noise
$N(\mu_I,\sigma_I^2)$ and $\tau_s < 1$ then the response function of the
neuron in equilibrium is given by  
\begin{equation}
\frac{1}{p(X)} = r + \sqrt{\pi}\int_{\frac{V_r-\mu_I - JX/N}{\sigma_I^2}}^{\frac{V_\theta-\mu_I - JX/N}{\sigma_I^2}} du e^{u^2}(1+erf(u))
\end{equation}
The mean activity is again given by $Nq$ where $\langle p(X)
\rangle_\mu=q$, and $q$ is a solution of (\ref{eq:fp}):
\begin{equation}
q = \left[ r + \sqrt{\pi}\int_{\frac{V_r-\mu_I -
      Jq}{\sigma_I^2}}^{\frac{V_\theta-\mu_I - Jq}{\sigma_I^2}} du
  e^{u^2}(1+erf(u))\right]^{-1}
\label{eq:qif}
\end{equation}
Using (\ref{eq:qif}), we can derive the slope factor $\lambda$,
\begin{equation}
\lambda = m^2\frac{J}{\sigma_I^2}\left(e^{(\frac{V_\theta-\mu_I - Jm}{\sigma_I^2})^2}(1+erf(\frac{V_\theta-\mu_I - Jm}{\sigma_I^2})-e^{(\frac{V_r-\mu_I - Jm}{\sigma_I^2})^2}(1+erf(\frac{V_r-\mu_I - Jm}{\sigma_I^2}))\right)
\label{eq:brunel_lambda}
\end{equation}
which can be used to estimate the variance using equation
(\ref{eq:finalvariance}).  

We also evaluate the response function numerically. We assume that
the response function can be approximated by the firing probability of a neuron 
whose membrane potential obeys
\begin{equation}
\tau \frac{dV}{dt}= I+\frac{JX}{N} -V(t) + Z(t)
\label{dV}
\end{equation}
where $X \in \{0,...,N\}$.  This presumes that the effect
of discrete synaptic events can be approximated by a constant input
equivalent to the mean input plus noise.
We estimated the firing probability in a temporal epoch from
the average firing rate of the neuron obeying (\ref{dV}) for each $X$.
We used the firing probabilities directly to form the MTM.  We then
computed the equilibrium PDF, the mean activity
(crossing point),  the slope factor, and the variance.  We choose the
bin size of the Markov model to be the membrane
time constant.  All differential equations were computed using the Euler method
with a time step of 0.01 ms.  

\begin{figure}[htbp]
\begin{center}
\includegraphics[width=4.25in]{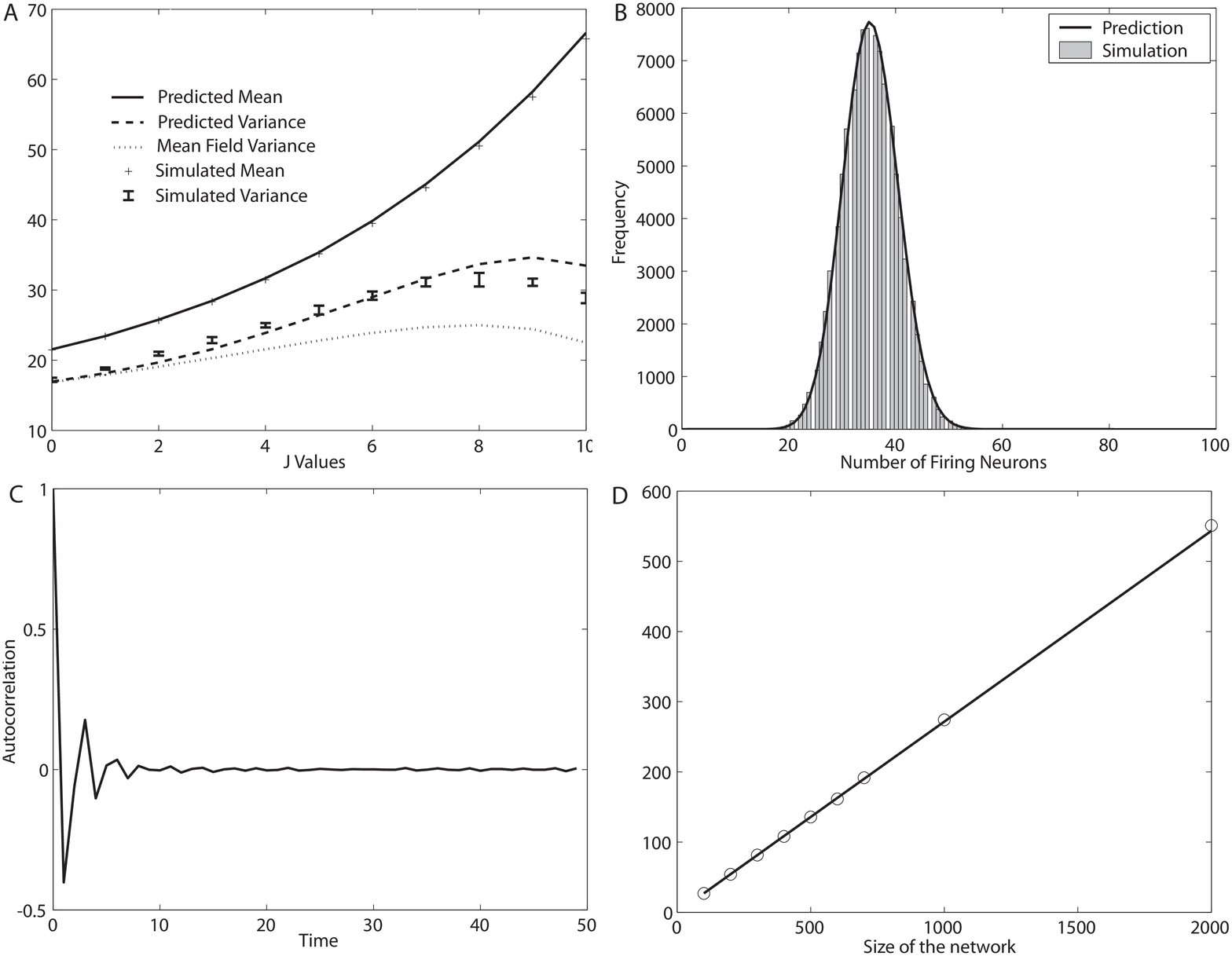}
\end{center}
\caption{Current Synapse model -- A) Mean (crosses) and variance (diamonds) versus $J$ from the numerical simulation compare well with the predictions (solid and dashed lines). The mean-field variance is also plotted (dotted line). Parameters are $N=100$, $I=4 $, $\sigma = 2.0$. B) The numerical PDF(histogram) and prediction (solid line) for $N=100$, $J=5.0$, $I=4 $, and $\sigma = 2.0$. C) The autocorrelation (solid line) for a network of $N=1000$. The dashed line is the estimated exponential decay with $\lambda=Cov(1)/Var$ (see text). Parameters are $J=5$, $I=4 $, $\sigma = 2.0$ and $\lambda=0.40$. D) The variance (circles) versus $N$. The solid line is from equation (\ref{eq:finalvariance}) with $\lambda$ computed using the autocorrelation decay for $N=1000$ ($\lambda=0.4$) displayed in C). Parameters are $J=5$, $I=4$, $\sigma = 2.0$.}
\label{fig:current}
\end{figure}

We compared the numerically simulated mean and variance of the network activity
at equilibrium for 100 neurons with our theoretical predictions
for differing synaptic strength $J$. The results are displayed
in figure \ref{fig:current}A. A close match is observed for both
quantities. The mean and variance increase with $J$.   Figure \ref{fig:current}A also shows the mean-field (uncorrelated) variance.  We see that the network variance always exceeds the mean-field value especially for midrange values of $J$.
%
%[For example, for $J=5$,
%the mean activity is 35.16 and the variance is 27.12
%The mean field variance should then be 22.79 which
%is lower, the ratio is then 1.18. This difference comes from the
%correlations. When $J=0$, the mean, variance and bernoulli
%(mean-field) variance are respectively 21.4, 17.2 and 16.8, the ratio
%falling to 1.02. The variability of the network in that case is
%entirely due to the noisy input. ]
The PDFs from the numerical simulation of 100 neurons and using the 
Markov model are shown in \ref{fig:current}B.  The simulated PDF was generated by taking a histogram of the network activity over $10^5$ ms.  For the Markov model, the PDF is the
eigenvector with eigenvalue one and approximated by taking
a single row of  the one hundredth power of the MTM.

The model predicts an exponentially decaying autocorrelation function
with  time constant $1/\ln(\lambda)$. It is displayed in figure
\ref{fig:current}C for a network of $N=1000$. We used the same
parameters as in figure \ref{fig:current}A with a connection weight where the dynamics
deviates from mean-field theory ($J=5$).  In this parameter range, the
recurrent excitation is strong and the single neuron firing rate is
very high, approximately 350 Hz (the refractory time of 1 ms
imposes a maximum frequency is 1000 Hz).   In this regime, the
refractory time acts like inhibition, so the probability to fire
actually decreases if the activity in the previous epoch increases.
Thus, the autocorrelation function exhibits anti-correlations as seen
in the figure \ref{fig:current}C.  We can estimate $\lambda$ by
fitting the autocorrelation to $\lambda^\tau$.  Using the
approximation $\lambda=Cov(1)/Var$, which gives $|\lambda|=0.40$.  The
theoretical value of $|\lambda|$ using (\ref{eq:brunel_lambda})
gives 0.37 for the same parameters (using the simulated mean number of
firing neurons).  Figure \ref{fig:current}D shows a plot of the
variance versus $N$ for the numerical simulation.  The simulated variance is
well matched by the estimated variance from (\ref{eq:finalvariance})
with $q$ measured from the simulation (estimated by the mean firing
rate of a single neuron) and $\lambda=0.4$.

\subsection{Conductance-based synapse}

We now consider the integrate-and-fire neuron model with conductance-based synaptic connections.
The membrane potential $V$ obeys
\begin{equation}
\label{eq:if}
\tau \frac{dV}{dt} = I(t) - (V-V_r) - s(t)\left(V - V_{s}\right) + Z(t)
\end{equation}
where $\tau$ is the membrane time constant, $I(t)$ is an input current, Z(t) is zero-mean white noise with variance $\sigma$,
$V_r$ is the rest potential, and $V_{s}$ is the reversal potential of an
excitatory synapse.  The synaptic gating variable $s(t)$ obeys
\begin{equation}
\label{eq:synapse}
\tau_s\frac{ds}{dt} =  \frac{J}{N} \sum_{t^f}\delta(t-t^f) -s(t)
\end{equation}
where  $\tau_s$ is the synaptic time constant, $J$ is the maximal synaptic conductance,
$N$ is the number of neurons in the network, and  $t^f$ are the
firing times of all neurons in the network. The threshold is $V_\theta$ and the
reset potential is $V_{R}$. As with the previous model, a refractory period $r$ was introduced. We used $\tau = 1$ ms, $\tau_s= 1$ ms,  $V_r
= -65$ mV, $V_s=0$ mV, $V_\theta = -60$ mV, $V_{R}=-80$ mV, 
$\tau_s = 1$ ms and $r=1$ ms.

\begin{figure}[htbp]
\begin{center}
\includegraphics[width=4.25in]{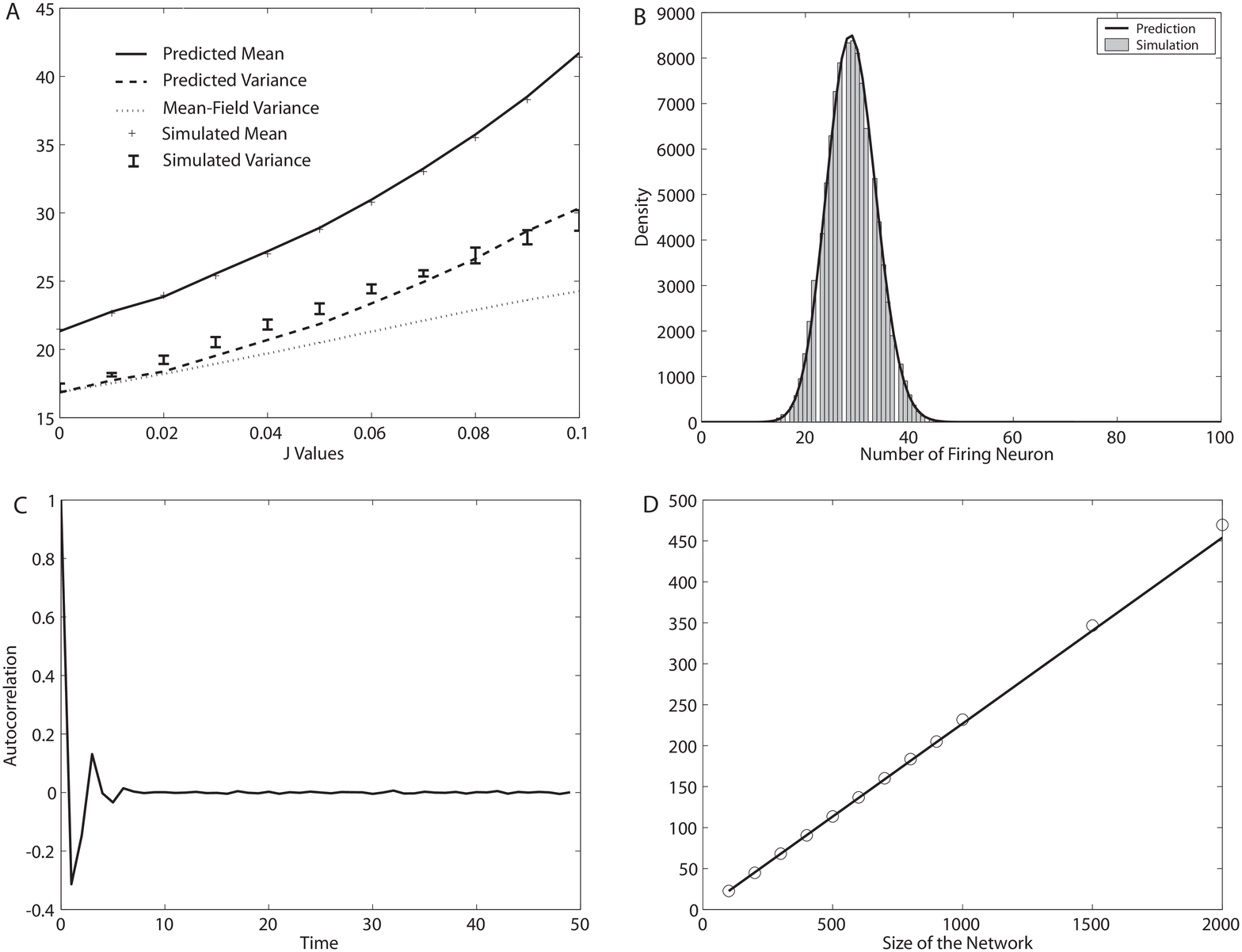}
\end{center}
\caption{Conductance Synapse model -- A) Mean (crosses) and variance (diamonds) versus $J$ from the numerical simulation compare well with the predictions (solid and dashed lines).  Parameters are $N=100$, $I=4 $, $\sigma = 2.0$.  B) The numerical PDF(histogram) and prediction (solid line) for $N=100$, $I=4 $, $\sigma = 2.0$, $J=0.05$. C) The autocorrelation (solid line) for a network of $N=1500$. The dashed line is the estimated exponential decay with $\lambda=Cov(1)/Var$ (see text). Parameters are $J=0.05$, $I=4 $, $\sigma = 2.0$ and $\lambda=0.31$. D) The variance (circles) versus $N$. The solid line is from equation (\ref{eq:finalvariance}) with $\lambda$ computed using the autocorrelation decay for $N=1500$ ($\lambda=0.31$) displayed in C). Parameters are $J=0.05$, $I=4$, $\sigma = 2.0$.}
\label{fig:conductance}
\end{figure}

We computed the response function by measuring numerically the firing probability of
a neuron that obeys 
\begin{equation}
\label{eq:resp_cond}
\tau \frac{dV}{dt} = I - (V-V_r) - \frac{JX}{N}\left(V - V_{rvs}\right) + Z(t)
\end{equation} 
for all $X \in \{0,...,N\}$ using the same method as in the current-based model.  From the response function, we obtained the MTM and the
invariant measure. Figure \ref{fig:conductance}A compares the mean and variance of a numerical simulation with the Markov prediction for
varying $J$. We see that there is a good match. For $J = 0.05$, we compare the numerically computed PDF with the invariant measure. As shown in figure
\ref{fig:conductance}B, the prediction is extremely accurate. We estimated the slope factor as in the previous section using the autocorrelation function for $N=1500$ and found $\lambda=0.31$ and computed the estimated variance for various $N$. The comparison is shown on figure \ref{fig:conductance}D. There is again a close match. 

\section{Discussion}

%By assuming statistical homogeneity and finite memory, we show that a
%complete statistical description of the neural activity of a
%finite-sized network can be determined.  
Our model relies on the assumption that the neuronal
dynamics can be represented by a Markov process.  We partition time
into discrete epochs and the probability that a neuron will fire in
one epoch only depends on the activity of the previous epoch.  For
this approximation to be valid, at least two conditions must be met.
The first is that the epoch time needs to be long enough so that the
influence of activity in the distant past is negligible.  The second
is that the epoch time needs to be short enough so that a neuron fires
at most once within an epoch.  The influence time of a neuron is given
approximately by the larger of the membrane time constant ($\tau_m$)
and synaptic time constant ($\tau_s$).  Presumably, the neuron does
not have a memory of events much beyond this time scale.  This gives a
lower bound on the epoch time $\Delta t$.  Hence, $\Delta t >
\max[\tau_m,\tau_s]$.
The second condition is equivalent to $f\Delta t<1$, where $f$ is the
max firing rate of a neuron in the network. Thus, the Markov model is
applicable for
\begin{equation}
\max[\tau_m,\tau_s] < f^{-1}
\end{equation}
This condition is not too difficult to satisfy.
For example, a cortical neuron receiving AMPA inputs with a membrane time constant less than 20 ms can satisfy the Markov condition if its maximum firing rate is below 50 Hz.   Since typical cortical neurons have a membrane time constant between 1 and
20 ms and many recordings in the cortex find rates below 50 Hz \cite{nicholls1992}, our formalism could be applicable to 
a broad class of cortical circuits.

The equilibrium state of our Markov model exists and is exactly
solvable if the response function is never zero or one.  In other
words,  a neuron in the network always has a nonzero probability to
fire but never has full certainty it will fire.  Hence, the neuronal
dynamics are always fully stochastic.  Thus the equilibrium state of a
fully stochastic network could be another definition of the
asynchronous state.
However, even though the network is completely stochastic, the activity is not
uncorrelated.  These correlations are manifested in the
entire network activity but not within the firing statistics of the
individual neurons, which obey a simple random
binomial or Poisson distribution.  The autocorrelation function of the individual
neuron also decays much more quickly than that of the network. 

Many previous approaches to studying the asynchronous state assumed that
neuronal firing was statistically independent so a mean-field
description was valid.  With our simple model, we can compute the
equilibrium PDF directly and
explicitly
determine the parameter regimes where mean-field theory breaks down.  
We also show that the ``order parameter'' that determines nearness to
criticality is the slope of the response function around the
equilibrium mean activity.  As expected, we find that mean-field
theory is less applicable for
small or strongly recurrent neural circuits.  In our model,
the mean activity of our network can be obtained using mean-field
theory except at the critical point.
We compute the variance of the network activity directly and see precisely how the
network dynamics deviate from mean-field theory as we approach
the critical point.  We note that
while our closed-form estimates for the mean and variance may break
down very
near criticality, our model does not.  The invariant measure of the
MTM still gives the equilibrium PDF and in principle can be computed
to arbitrary precision.  Hence, our model can serve as a simple
example of when mean-field theory is
applicable. We note that
even in the limit of $N$ going to infinity, the variance of the
network deviates from a network of independent neurons by a factor of
$(1-\lambda)^{-1}$.  The deviations from mean-field theory persist
even in an infinite 
sized network.
 
Our model requires statistical homogeneity.  While purely homogeneous
networks are unrealistic for biological networks,
statistical homogeneity may be more plausible. For some classes of
inputs, the probability of firing for a given set of neurons could be
uniform over some limited temporal duration.  We found that
our analytical estimates can be modified to account for disordered
randomness in the connections.
A model that fully incorporated all heterogeneous effects would require
taking into account different populations.   
However, with each additional population, the
dimension of the MTM increases by a power of $N$.  Thus for a network
of two populations, say inhibitory and excitatory neurons, the resulting  problem will
involve a MTM with $N^2\times N^2$ elements. 

Future work could examine the behavior at the critical point more
carefully.  Our variance estimate predicted that the variance at
criticality would
diverge as the system size squared but simulations found an exponent
of 3/2.  We also showed that the correlation time of the network activity
diverged at the critical 
point.  We expect correlations to obey a power law at
criticality. Perhaps, a renormalization group approach may be adapted
to study the scaling behavior near criticality.
Recent experiments
have found that cortical slices exhibit critical
behavior~\cite{beggs2004,beggs2003}.  Our Markov model may be
an ideal system to explore critical behavior.

Finally, we note that
even at the critical point, where the network activity is highly
correlated, the single neuron dynamics can exhibit uncorrelated Poisson
statistics. 
This shows that collective network dynamics may not be deducible from
single neuron dynamics.  Using our model as a basis, it may be
possible to probe characteristics about local network size,
connectivity and nearness to criticality by combining recordings of
single neurons with measurements of the local field potential.

\section{Acknowledgments}
This works was supported by the Intramural Research Program of NIH, NIDDK. We would like to thank Michael Buice for his helpful comments on the manuscript.

\bibliography{soulachow}

\begin{thebibliography}{42}
\expandafter\ifx\csname natexlab\endcsname\relax\def\natexlab#1{#1}\fi
\expandafter\ifx\csname bibnamefont\endcsname\relax
  \def\bibnamefont#1{#1}\fi
\expandafter\ifx\csname bibfnamefont\endcsname\relax
  \def\bibfnamefont#1{#1}\fi
\expandafter\ifx\csname citenamefont\endcsname\relax
  \def\citenamefont#1{#1}\fi
\expandafter\ifx\csname url\endcsname\relax
  \def\url#1{\texttt{#1}}\fi
\expandafter\ifx\csname urlprefix\endcsname\relax\def\urlprefix{URL }\fi
\providecommand{\bibinfo}[2]{#2}
\providecommand{\eprint}[2][]{\url{#2}}

\bibitem[{\citenamefont{Singer and Gray}(1995)}]{singer1995}
\bibinfo{author}{\bibfnamefont{W.}~\bibnamefont{Singer}} \bibnamefont{and}
  \bibinfo{author}{\bibfnamefont{C.~M.} \bibnamefont{Gray}},
  \bibinfo{journal}{Annu Rev Neurosci} \textbf{\bibinfo{volume}{18}},
  \bibinfo{pages}{555} (\bibinfo{year}{1995}).

\bibitem[{\citenamefont{Fries et~al.}(2001)\citenamefont{Fries, Reynolds,
  Rorie, and Desimone}}]{fries2001}
\bibinfo{author}{\bibfnamefont{P.}~\bibnamefont{Fries}},
  \bibinfo{author}{\bibfnamefont{J.~H.} \bibnamefont{Reynolds}},
  \bibinfo{author}{\bibfnamefont{A.~E.} \bibnamefont{Rorie}}, \bibnamefont{and}
  \bibinfo{author}{\bibfnamefont{R.}~\bibnamefont{Desimone}},
  \bibinfo{journal}{Science} \textbf{\bibinfo{volume}{291}},
  \bibinfo{pages}{1560} (\bibinfo{year}{2001}).

\bibitem[{\citenamefont{Pesaran et~al.}(2002)\citenamefont{Pesaran, Pezaris,
  Sahani, Mitra, and Andersen}}]{pesaran2002}
\bibinfo{author}{\bibfnamefont{B.}~\bibnamefont{Pesaran}},
  \bibinfo{author}{\bibfnamefont{J.~S.} \bibnamefont{Pezaris}},
  \bibinfo{author}{\bibfnamefont{M.}~\bibnamefont{Sahani}},
  \bibinfo{author}{\bibfnamefont{P.~P.} \bibnamefont{Mitra}}, \bibnamefont{and}
  \bibinfo{author}{\bibfnamefont{R.~A.} \bibnamefont{Andersen}},
  \bibinfo{journal}{Nat Neurosci} \textbf{\bibinfo{volume}{5}},
  \bibinfo{pages}{805} (\bibinfo{year}{2002}).

\bibitem[{\citenamefont{Steinmetz et~al.}(2000)\citenamefont{Steinmetz, Roy,
  Fitzgerald, Hsiao, Johnson, and Niebur}}]{steinmetz2000}
\bibinfo{author}{\bibfnamefont{P.~N.} \bibnamefont{Steinmetz}},
  \bibinfo{author}{\bibfnamefont{A.}~\bibnamefont{Roy}},
  \bibinfo{author}{\bibfnamefont{P.~J.} \bibnamefont{Fitzgerald}},
  \bibinfo{author}{\bibfnamefont{S.~S.} \bibnamefont{Hsiao}},
  \bibinfo{author}{\bibfnamefont{K.~O.} \bibnamefont{Johnson}},
  \bibnamefont{and} \bibinfo{author}{\bibfnamefont{E.}~\bibnamefont{Niebur}},
  \bibinfo{journal}{Nature} \textbf{\bibinfo{volume}{404}},
  \bibinfo{pages}{187} (\bibinfo{year}{2000}).

\bibitem[{\citenamefont{Softky and Koch}(1993)}]{softky1993}
\bibinfo{author}{\bibfnamefont{W.~R.} \bibnamefont{Softky}} \bibnamefont{and}
  \bibinfo{author}{\bibfnamefont{C.}~\bibnamefont{Koch}}, \bibinfo{journal}{J
  Neurosci} \textbf{\bibinfo{volume}{13}}, \bibinfo{pages}{334}
  (\bibinfo{year}{1993}).

\bibitem[{\citenamefont{Amit and Brunel}(1997a)}]{amit1997a}
\bibinfo{author}{\bibfnamefont{D.~J.} \bibnamefont{Amit}} \bibnamefont{and}
  \bibinfo{author}{\bibfnamefont{N.}~\bibnamefont{Brunel}},
  \bibinfo{journal}{Network: Comput. Neural. Syst.}
  \textbf{\bibinfo{volume}{8}}, \bibinfo{pages}{373} (\bibinfo{year}{1997a}).

\bibitem[{\citenamefont{Amit and Brunel}(1997b)}]{amit1997b}
\bibinfo{author}{\bibfnamefont{D.~J.} \bibnamefont{Amit}} \bibnamefont{and}
  \bibinfo{author}{\bibfnamefont{N.}~\bibnamefont{Brunel}},
  \bibinfo{journal}{Cereb Cortex} \textbf{\bibinfo{volume}{7}},
  \bibinfo{pages}{237} (\bibinfo{year}{1997b}).

\bibitem[{\citenamefont{Van~Vreeswijk and
  Sompolinsky}(1996)}]{vanvreeswijk1996}
\bibinfo{author}{\bibfnamefont{C.}~\bibnamefont{Van~Vreeswijk}}
  \bibnamefont{and}
  \bibinfo{author}{\bibfnamefont{H.}~\bibnamefont{Sompolinsky}},
  \bibinfo{journal}{Science} \textbf{\bibinfo{volume}{274}},
  \bibinfo{pages}{1724} (\bibinfo{year}{1996}).

\bibitem[{\citenamefont{Van~Vreeswijk and
  Sompolinsky}(1998)}]{vanvreeswijk1998}
\bibinfo{author}{\bibfnamefont{C.}~\bibnamefont{Van~Vreeswijk}}
  \bibnamefont{and}
  \bibinfo{author}{\bibfnamefont{H.}~\bibnamefont{Sompolinsky}},
  \bibinfo{journal}{Neural Comput} \textbf{\bibinfo{volume}{10}},
  \bibinfo{pages}{1321} (\bibinfo{year}{1998}).

\bibitem[{\citenamefont{Gerstner}(2000)}]{gerstner2000}
\bibinfo{author}{\bibfnamefont{W.}~\bibnamefont{Gerstner}},
  \bibinfo{journal}{Neural Computation} \textbf{\bibinfo{volume}{12}},
  \bibinfo{pages}{43} (\bibinfo{year}{2000}).

\bibitem[{\citenamefont{Cai et~al.}(2004)\citenamefont{Cai, Tao, Shelley, and
  McLaughlin}}]{cai2004}
\bibinfo{author}{\bibfnamefont{D.}~\bibnamefont{Cai}},
  \bibinfo{author}{\bibfnamefont{L.}~\bibnamefont{Tao}},
  \bibinfo{author}{\bibfnamefont{M.}~\bibnamefont{Shelley}}, \bibnamefont{and}
  \bibinfo{author}{\bibfnamefont{D.~W.} \bibnamefont{McLaughlin}},
  \bibinfo{journal}{Proc Natl Acad Sci U S A} \textbf{\bibinfo{volume}{101}},
  \bibinfo{pages}{7757} (\bibinfo{year}{2004}).

\bibitem[{\citenamefont{Gerstner and Kistler}(2002a)}]{gerstner2002a}
\bibinfo{author}{\bibfnamefont{W.}~\bibnamefont{Gerstner}} \bibnamefont{and}
  \bibinfo{author}{\bibfnamefont{W.}~\bibnamefont{Kistler}},
  \emph{\bibinfo{title}{Spiking Neuron Models - Single Neurons, Populations,
  Plasticity}} (\bibinfo{publisher}{Cambrige University Press},
  \bibinfo{address}{Cambridge, UK}, \bibinfo{year}{2002a}).

\bibitem[{\citenamefont{Abbott and Vreeswijk}(1993)}]{abbott1993}
\bibinfo{author}{\bibfnamefont{L.}~\bibnamefont{Abbott}} \bibnamefont{and}
  \bibinfo{author}{\bibfnamefont{C.}~\bibnamefont{Vreeswijk}},
  \bibinfo{journal}{Phys. Rev. E} \textbf{\bibinfo{volume}{48}},
  \bibinfo{pages}{1483} (\bibinfo{year}{1993}).

\bibitem[{\citenamefont{Treves}(1993)}]{treves1993}
\bibinfo{author}{\bibfnamefont{A.}~\bibnamefont{Treves}},
  \bibinfo{journal}{Network} \textbf{\bibinfo{volume}{4}}, \bibinfo{pages}{259}
  (\bibinfo{year}{1993}).

\bibitem[{\citenamefont{Fusi and Mattia}(1999)}]{fusi1999}
\bibinfo{author}{\bibfnamefont{S.}~\bibnamefont{Fusi}} \bibnamefont{and}
  \bibinfo{author}{\bibfnamefont{M.}~\bibnamefont{Mattia}},
  \bibinfo{journal}{Neural Computation} \textbf{\bibinfo{volume}{11}},
  \bibinfo{pages}{633} (\bibinfo{year}{1999}).

\bibitem[{\citenamefont{Golomb and Hansel}(2000)}]{golomb2000}
\bibinfo{author}{\bibfnamefont{D.}~\bibnamefont{Golomb}} \bibnamefont{and}
  \bibinfo{author}{\bibfnamefont{D.}~\bibnamefont{Hansel}},
  \bibinfo{journal}{Neural Comput} \textbf{\bibinfo{volume}{12}},
  \bibinfo{pages}{1095} (\bibinfo{year}{2000}).

\bibitem[{\citenamefont{Brunel}(2000a)}]{brunel2000a}
\bibinfo{author}{\bibfnamefont{N.}~\bibnamefont{Brunel}}, \bibinfo{journal}{J
  Comput Neurosci} \textbf{\bibinfo{volume}{8}}, \bibinfo{pages}{183}
  (\bibinfo{year}{2000a}).

\bibitem[{\citenamefont{Nykamp and Tranchina}(2000)}]{nykamp2000}
\bibinfo{author}{\bibfnamefont{D.}~\bibnamefont{Nykamp}} \bibnamefont{and}
  \bibinfo{author}{\bibfnamefont{D.}~\bibnamefont{Tranchina}},
  \bibinfo{journal}{Journal of Computational Neuroscience}
  \textbf{\bibinfo{volume}{8}}, \bibinfo{pages}{19} (\bibinfo{year}{2000}).

\bibitem[{\citenamefont{Hansel and Mato}(2003)}]{hansel2003}
\bibinfo{author}{\bibfnamefont{D.}~\bibnamefont{Hansel}} \bibnamefont{and}
  \bibinfo{author}{\bibfnamefont{G.}~\bibnamefont{Mato}},
  \bibinfo{journal}{Neural Comput} \textbf{\bibinfo{volume}{15}},
  \bibinfo{pages}{1} (\bibinfo{year}{2003}).

\bibitem[{\citenamefont{Del~Giudice and Mattia}(2003)}]{delgiudice2003}
\bibinfo{author}{\bibfnamefont{P.}~\bibnamefont{Del~Giudice}} \bibnamefont{and}
  \bibinfo{author}{\bibfnamefont{M.}~\bibnamefont{Mattia}}, in
  \emph{\bibinfo{booktitle}{Advances in Condensed Matter and Statistical
  Physics}} (\bibinfo{publisher}{Nova Science}, \bibinfo{address}{Hauppauge,
  NY}, \bibinfo{year}{2003}), pp. \bibinfo{pages}{125--153}.

\bibitem[{\citenamefont{Brunel and Hansel}(2006)}]{brunel2006}
\bibinfo{author}{\bibfnamefont{N.}~\bibnamefont{Brunel}} \bibnamefont{and}
  \bibinfo{author}{\bibfnamefont{D.}~\bibnamefont{Hansel}},
  \bibinfo{journal}{Neural Comput} \textbf{\bibinfo{volume}{18}},
  \bibinfo{pages}{1066} (\bibinfo{year}{2006}).

\bibitem[{\citenamefont{Plesser and Gerstner}(2000)}]{plesser2000}
\bibinfo{author}{\bibfnamefont{H.~E.} \bibnamefont{Plesser}} \bibnamefont{and}
  \bibinfo{author}{\bibfnamefont{W.}~\bibnamefont{Gerstner}},
  \bibinfo{journal}{Neural Computation} \textbf{\bibinfo{volume}{12}},
  \bibinfo{pages}{367} (\bibinfo{year}{2000}).

\bibitem[{\citenamefont{Salinas and Sejnowski}(2002)}]{salinas2002}
\bibinfo{author}{\bibfnamefont{E.}~\bibnamefont{Salinas}} \bibnamefont{and}
  \bibinfo{author}{\bibfnamefont{T.~J.} \bibnamefont{Sejnowski}},
  \bibinfo{journal}{Neural Comput} \textbf{\bibinfo{volume}{14}},
  \bibinfo{pages}{2111} (\bibinfo{year}{2002}).

\bibitem[{\citenamefont{Fourcaud and Brunel}(2002)}]{fourcaud2002}
\bibinfo{author}{\bibfnamefont{N.}~\bibnamefont{Fourcaud}} \bibnamefont{and}
  \bibinfo{author}{\bibfnamefont{N.}~\bibnamefont{Brunel}},
  \bibinfo{journal}{Neural Comput} \textbf{\bibinfo{volume}{14}},
  \bibinfo{pages}{2057} (\bibinfo{year}{2002}).

\bibitem[{\citenamefont{Soula et~al.}(2006)\citenamefont{Soula, Beslon, and
  Mazet}}]{soula2006}
\bibinfo{author}{\bibfnamefont{H.}~\bibnamefont{Soula}},
  \bibinfo{author}{\bibfnamefont{G.}~\bibnamefont{Beslon}}, \bibnamefont{and}
  \bibinfo{author}{\bibfnamefont{O.}~\bibnamefont{Mazet}},
  \bibinfo{journal}{Neural Computation} \textbf{\bibinfo{volume}{18}},
  \bibinfo{pages}{60} (\bibinfo{year}{2006}).

\bibitem[{\citenamefont{Ginzburg and Sompolinsky}(1994)}]{ginzburg1994}
\bibinfo{author}{\bibfnamefont{I.}~\bibnamefont{Ginzburg}} \bibnamefont{and}
  \bibinfo{author}{\bibfnamefont{H.}~\bibnamefont{Sompolinsky}},
  \bibinfo{journal}{Phys Rev E Stat Nonlin Soft Matter Phys}
  \textbf{\bibinfo{volume}{50}}, \bibinfo{pages}{3171} (\bibinfo{year}{1994}).

\bibitem[{\citenamefont{Meyer and Van~Vreeswijk}(2002)}]{meyer2002}
\bibinfo{author}{\bibfnamefont{C.}~\bibnamefont{Meyer}} \bibnamefont{and}
  \bibinfo{author}{\bibfnamefont{C.}~\bibnamefont{Van~Vreeswijk}},
  \bibinfo{journal}{Neural Computation} \textbf{\bibinfo{volume}{14}},
  \bibinfo{pages}{369} (\bibinfo{year}{2002}).

\bibitem[{\citenamefont{Brunel}(2003b)}]{brunel2003b}
\bibinfo{author}{\bibfnamefont{N.}~\bibnamefont{Brunel}},
  \bibinfo{journal}{Cereb Cortex} \textbf{\bibinfo{volume}{13}},
  \bibinfo{pages}{1151} (\bibinfo{year}{2003b}).

\bibitem[{\citenamefont{Mattia and del Giudice}(2002)}]{mattia2002}
\bibinfo{author}{\bibfnamefont{M.}~\bibnamefont{Mattia}} \bibnamefont{and}
  \bibinfo{author}{\bibfnamefont{P.}~\bibnamefont{del Giudice}},
  \bibinfo{journal}{Physical Review E} \textbf{\bibinfo{volume}{66}}
  (\bibinfo{year}{2002}).

\bibitem[{\citenamefont{Gerstner and Hemmen}(1992)}]{gerstner1992}
\bibinfo{author}{\bibfnamefont{W.}~\bibnamefont{Gerstner}} \bibnamefont{and}
  \bibinfo{author}{\bibfnamefont{J.~v.} \bibnamefont{Hemmen}},
  \bibinfo{journal}{Network} \textbf{\bibinfo{volume}{3}}, \bibinfo{pages}{139}
  (\bibinfo{year}{1992}).

\bibitem[{\citenamefont{Gerstner}(1995)}]{gerstner1995}
\bibinfo{author}{\bibfnamefont{W.}~\bibnamefont{Gerstner}},
  \bibinfo{journal}{Physical Review. E. Statistical Physics, Plasmas, Fluids,
  and Related Interdisciplinary Topics} \textbf{\bibinfo{volume}{51}},
  \bibinfo{pages}{738} (\bibinfo{year}{1995}).

\bibitem[{\citenamefont{Seung}(1996)}]{seung1996}
\bibinfo{author}{\bibfnamefont{H.~S.} \bibnamefont{Seung}},
  \bibinfo{journal}{Proc Natl Acad Sci U S A} \textbf{\bibinfo{volume}{93}},
  \bibinfo{pages}{13339} (\bibinfo{year}{1996}).

\bibitem[{\citenamefont{Kuznetsov}(1998)}]{kuznetsov1998}
\bibinfo{author}{\bibfnamefont{Y.}~\bibnamefont{Kuznetsov}},
  \emph{\bibinfo{title}{Elements of Bifurcation Theory 2nd ed.}}
  (\bibinfo{publisher}{Springer}, \bibinfo{year}{1998}).

\bibitem[{\citenamefont{Compte et~al.}(2000)\citenamefont{Compte, Brunel,
  Goldman-Rakic, and Wang}}]{compte2000}
\bibinfo{author}{\bibfnamefont{A.}~\bibnamefont{Compte}},
  \bibinfo{author}{\bibfnamefont{N.}~\bibnamefont{Brunel}},
  \bibinfo{author}{\bibfnamefont{P.~S.} \bibnamefont{Goldman-Rakic}},
  \bibnamefont{and} \bibinfo{author}{\bibfnamefont{X.~J.} \bibnamefont{Wang}},
  \bibinfo{journal}{Cereb Cortex} \textbf{\bibinfo{volume}{10}},
  \bibinfo{pages}{910} (\bibinfo{year}{2000}).

\bibitem[{\citenamefont{Laing and Chow}(2001)}]{laing2001}
\bibinfo{author}{\bibfnamefont{C.~R.} \bibnamefont{Laing}} \bibnamefont{and}
  \bibinfo{author}{\bibfnamefont{C.~C.} \bibnamefont{Chow}},
  \bibinfo{journal}{Neural Comput} \textbf{\bibinfo{volume}{13}},
  \bibinfo{pages}{1473} (\bibinfo{year}{2001}).

\bibitem[{\citenamefont{Gutkin et~al.}(2001)\citenamefont{Gutkin, Laing, Colby,
  Chow, and Ermentrout}}]{gutkin2001}
\bibinfo{author}{\bibfnamefont{B.~S.} \bibnamefont{Gutkin}},
  \bibinfo{author}{\bibfnamefont{C.~R.} \bibnamefont{Laing}},
  \bibinfo{author}{\bibfnamefont{C.~L.} \bibnamefont{Colby}},
  \bibinfo{author}{\bibfnamefont{C.~C.} \bibnamefont{Chow}}, \bibnamefont{and}
  \bibinfo{author}{\bibfnamefont{G.~B.} \bibnamefont{Ermentrout}},
  \bibinfo{journal}{J Comput Neurosci} \textbf{\bibinfo{volume}{11}},
  \bibinfo{pages}{121} (\bibinfo{year}{2001}).

\bibitem[{\citenamefont{Compte}(2006)}]{compte2006}
\bibinfo{author}{\bibfnamefont{A.}~\bibnamefont{Compte}},
  \bibinfo{journal}{Neuroscience} \textbf{\bibinfo{volume}{139}},
  \bibinfo{pages}{135} (\bibinfo{year}{2006}).

\bibitem[{\citenamefont{Risken}(1989)}]{risken1989}
\bibinfo{author}{\bibfnamefont{H.}~\bibnamefont{Risken}},
  \emph{\bibinfo{title}{The Fokker-Planck Equation: Methods of Solution and
  Application 2nd ed}} (\bibinfo{publisher}{Springer}, \bibinfo{year}{1989}).

\bibitem[{\citenamefont{Hopfield}(1982)}]{hopfield1982}
\bibinfo{author}{\bibfnamefont{J.}~\bibnamefont{Hopfield}},
  \bibinfo{journal}{Proc. Natl. Acad. Sci. USA} \textbf{\bibinfo{volume}{79}},
  \bibinfo{pages}{2554} (\bibinfo{year}{1982}).

\bibitem[{\citenamefont{Nicholls et~al.}(1992)\citenamefont{Nicholls, Martin,
  and Wallace}}]{nicholls1992}
\bibinfo{author}{\bibfnamefont{J.}~\bibnamefont{Nicholls}},
  \bibinfo{author}{\bibfnamefont{A.}~\bibnamefont{Martin}}, \bibnamefont{and}
  \bibinfo{author}{\bibfnamefont{B.}~\bibnamefont{Wallace}},
  \emph{\bibinfo{title}{From Neuron To Brain Third Ed.}}
  (\bibinfo{publisher}{Sinauer Associates, Inc}, \bibinfo{address}{Sunderland,
  MA}, \bibinfo{year}{1992}).

\bibitem[{\citenamefont{Beggs and Plenz}(2004)}]{beggs2004}
\bibinfo{author}{\bibfnamefont{J.~M.} \bibnamefont{Beggs}} \bibnamefont{and}
  \bibinfo{author}{\bibfnamefont{D.}~\bibnamefont{Plenz}}, \bibinfo{journal}{J
  Neurosci} \textbf{\bibinfo{volume}{24}}, \bibinfo{pages}{5216}
  (\bibinfo{year}{2004}).

\bibitem[{\citenamefont{Beggs and Plenz}(2003)}]{beggs2003}
\bibinfo{author}{\bibfnamefont{J.~M.} \bibnamefont{Beggs}} \bibnamefont{and}
  \bibinfo{author}{\bibfnamefont{D.}~\bibnamefont{Plenz}}, \bibinfo{journal}{J
  Neurosci} \textbf{\bibinfo{volume}{23}}, \bibinfo{pages}{11167}
  (\bibinfo{year}{2003}).

\end{thebibliography}

\begin{appendix}
\appendix
\section{Mean and variance derivations}
\label{app:mean}
The mean activity is given by
\begin{eqnarray*}
\langle X \rangle _t &=& \sum_{k=0}^N kP_k(t) \\
&=& \sum_{k=0}^N k \sum_{i=0}^N C_N^k (1-p_i)^{N-k}p_i^k P_i(t-1)
\end{eqnarray*}
Rearranging yields
$$
 \langle X \rangle _t =\sum_{i=0}^N P_i(t-1) \sum_{k=0}^N kC_N^k (1-p_i)^{N-k}p_i^k 
$$
 Since $\sum_{k=0}^N kC_N^k (1-p_i)^{N-k}p_i^k=Np_i$ is the mean of a binomial distribution, we obtain
 \begin{eqnarray*}
\langle X \rangle _t&=& \sum_{i=0}^N P_i(t-1) Np_i \\
&=& N\langle p \rangle _{t-1}
\end{eqnarray*}
which is (\ref{eq:markov_mean}).

The variance is given by
\begin{eqnarray*}
Var_t(X) &=& \sum_{k=0}^N k^2P_k(t) - N^2\langle p \rangle _{t-1}^2\\
&=& \sum_{k=0}^N k^2 \sum_{i=0}^N C_N^k (1-p_i)^{N-k}p_i^k P_i(t-1)- N^2\langle p \rangle _{t-1}^2\\
&=& \sum_{i=0}^N P_i(t-1) \sum_{k=0}^N k^2C_N^k (1-p_i)^{N-k}p_i^k - N^2\langle p \rangle _{t-1}^2\\
&=& \sum_{i=0}^N P_i(t-1) Np_i(1-p_i) + \sum_{i=0}^N P_i(t-1) N^2p_i^2  - N^2\langle p \rangle _{t-1}^2\\
&=& N\langle p(1-p) \rangle _{t-1} + N^2Var_{t-1}(p)
\end{eqnarray*}
where we have used the variance of a binomial distribution $Np(1-p)$. For the linear case, writing $p(X) = a+bX$ we have
\begin{eqnarray*}
Var_\mu(X) &=& N\langle  p(1-p) \rangle _\mu + N^2\langle p^2 \rangle _\mu - N^2\langle p \rangle _\mu^2\\
&=& N\langle a+bX-a^2-2abX-b^2X^2 \rangle _\mu + N^2\langle a^2+2abX+b^2X^2 \rangle _\mu  - N^2\langle a+bX \rangle _\mu^2\\
&=& Na +bN\langle X \rangle  -Na^2 -2Nab\langle X \rangle  - Nb^2\langle X^2 \rangle  +N^2a^2 + 2N^2ab\langle X \rangle  + N^2b^2\langle X^2 \rangle \\
& & - N^2(a^2 + 2ab\langle X \rangle  +b^2\langle X \rangle ^2)\\
&=& Na-Na^2 + N^2a^2 - N^2a^2 +\langle X \rangle (bN -2Nab +2N^2ab -2N^2ab)\\
& &  + \langle X^2 \rangle (-Nb^2+ N^2b^2) -b^2N^2\langle X \rangle ^2\\
&=& Na-Na^2 + bN(1-2a)\langle X \rangle  + (N^2b^2-Nb^2)Var(X) - Nb^2\langle X \rangle ^2\\
&=& \frac{Na-Na^2 + bN(1-2a)\langle X \rangle  - Nb^2\langle X \rangle ^2}{1 - N^2b^2 +Nb^2}
\end{eqnarray*}
Setting $a=p_0$ and $b=\frac{(q-p_0)}{Nm}$ gives
\begin{equation}
Var_\mu(X) =  q^2N\frac{p_0(1 -p_0)+(q-p_0)(1-p_0 -q)}{q^2 - (q-p_0)^2 + (q-p_0)^2/N }
\end{equation}
If we set $\lambda = (q-p_0)/q$ we get equation (\ref{eq:finalvariance}). 

\section{Autocovariance function}
\label{app:ac}
We prove the form of the autocovariance function $Cov(\tau)=\lambda^\tau Var_\mu(X)$ for the linear response function using induction.  We first show that $Cov(1)=\lambda Var_\mu(X)$ and then
 $Cov(\tau+1)=\lambda Cov(\tau)$.

The autocovariance function when $\tau=1$ is given by
\begin{eqnarray*}
\rm{Cov}(1)&=& \langle X_{t}X_{t+1}\rangle -\langle X\rangle^2_\mu \\
&=& \sum_{k=0}^N\sum_{i=0}^N kiP(X_{t+1}=i | X_t=k)P(X_t=k) -\langle X\rangle^2_\mu \\
&=& N\sum_{k=0}^N kp_kP(X_t=k) -\langle X\rangle^2_\mu \\
&=& N \langle pX \rangle_\mu -\langle X\rangle^2_\mu 
\end{eqnarray*}
For a linear  response function $p$, we obtain
\begin{eqnarray*}
\rm{Cov}(1)&=& N \langle pX \rangle_\mu -\langle X\rangle^2_\mu \\
&=&  N \langle p_0X + \frac{q-p_0}{Nq}X^2 \rangle_\mu -N^2m^2\\
&=& N^2p_0q + \langle \frac{q-p_0}{q}X^2\rangle_\mu -N^2q^2\\
&=& N^2q\left(p_0 -q\right) + \frac{q-p_0}{q}\left(Var_\mu(X) + N^2q^2\right)\\
&=& \frac{q-p_0}{q}Var_\mu(X)
\end{eqnarray*}
Hence for $\tau=1$, the autocovariance function is equal to the slope factor $\lambda = (q-p_0)/q$ times the variance.  Assume for $\tau$, that $Cov(\tau)=\lambda^\tau Var_\mu(X)$, then 
\begin{eqnarray*}
Cov(\tau+1) &=&\langle X_{t}X_{t+\tau+1}\rangle -\langle X\rangle^2_\mu \\ 
&=& \sum_{k=0}^{N}\sum_{j=0}^N kj P(X_t=k)P(X_{t+\tau+1}=j | X_t=k) -\langle X\rangle^2_\mu \\
&=& \sum_{k=0}^{N}\sum_{j=0}^N kj P(X_t=k)\sum_{r=0}^N P(X_{t+\tau+1}=j | X_{t+\tau}=r)P(X_{t+\tau}=r |X_t=k)-\langle X\rangle^2_\mu  \\
&=& \sum_{k=0}^{N}\sum_{j=0}^N \sum_{r=0}^N kj P(X_t=k)C_N^j (1-p(r))^{N-j}p(r)^j P(X_{t+\tau}=r |X_t=k)-\langle X\rangle^2_\mu \\
&=& \sum_{k=0}^{N}\sum_{r=0}^N Nkp(r) P(X_t=k)P(X_{t+\tau}=r |X_t=k)-\langle X\rangle^2_\mu 
\end{eqnarray*}
using the mean of the binomial distribution. We can now insert the
linear response function to obtain
\begin{eqnarray*}
Cov(\tau+1)&=& \sum_{k=0}^{N}\sum_{r=0}^N Nk(p_0 + \frac{q-p_0}{Nq}r) P(X_t=k)P(X_{t+\tau}=r |X_t=k) -\langle X\rangle^2_\mu \\
&=& \sum_{k=0}^{N}\sum_{r=0}^N Nkp_0P(X_t=k)P(X_{t+\tau}=r |X_t=k) + \frac{q-p_0}{q}\left(Cov(\tau)+\langle X\rangle^2_\mu\right) -\langle X\rangle^2_\mu 
\end{eqnarray*}
because $$\sum_{k=0}^{N}\sum_{r=0}^NkrP(X_t=k)P(X_{t+\tau}=r |X_t=k)= Cov(t+\tau)+ \langle X\rangle^2_\mu $$
Then since $$\sum_{k=0}^{N}\sum_{r=0}^N Nkp_0P(X_t=k)P(X_{t+\tau}=r |X_t=k)=\sum_{k=0}^{N}Nkp_0P(X_t=k)$$ because 
$$\sum_{r=0}^N P(X_{t+\tau}=r |X_t=k) =1$$ for all $k$ and  $$\sum_{k=0}^{N}Nkp_0P(X_t=k)=\langle Np_0X\rangle_\mu = N^2p_0q$$ we finally obtain
\begin{eqnarray*}
Cov(\tau+1)&=& N^2p_0q + \frac{q-p_0}{q}\langle X\rangle^2_\mu -\langle X\rangle^2_\mu + \lambda Cov(\tau)\\
&=& N^2p_0q + \frac{q-p_0}{q}N^2q^2 - N^2q^2 +  \lambda Cov(\tau)\\
&=& N^2(p_0q +(q-p_0)q -q^2) + \lambda Cov(\tau)\\
&=& \lambda Cov(\tau)
\end{eqnarray*}
proving equation (\ref{eq:autoone}) by induction. 

\end{appendix}

\newpage

\section*{Figures}

%\newpage

%\newpage

%\newpage

%\newpage

%\newpage

%\newpage
%\newpage

%\newpage

\end{document}